\documentclass[aps,prd,notitlepage,showpacs,nofootinbib,superscriptaddress]{revtex4-1}

\usepackage{graphicx}
\usepackage[utf8]{inputenc} 
\usepackage{amsmath}
\usepackage[T1]{fontenc}
\usepackage{amssymb}
\usepackage{float}
\usepackage{mathrsfs} 
\usepackage{comment}
\usepackage{slashed}
\usepackage[normalem]{ulem}
\usepackage{booktabs}
\usepackage{mathtools,braket,blkarray}
\usepackage[usenames,dvipsnames]{color}
\usepackage[colorinlistoftodos]{todonotes}
\usepackage[colorlinks=true,citecolor=darkred,urlcolor=darkred, pdfborder={0 0 0}]{hyperref}
\usepackage[normalem]{ulem}
\usepackage{multirow}
\usepackage{xfrac}
\usepackage[colorinlistoftodos]{todonotes}
\usepackage{soul}
\usepackage{subcaption}
\usepackage{ragged2e}
\usepackage{caption}
\DeclareCaptionFormat{justified}{\justifying#1#2#3}
\usepackage{rotating}

\newcommand {\ignore}[1]{}

\definecolor{linkcolor}{rgb}{0,0,0.5}
\definecolor{darkred}{rgb}{0.6,0,0}
\definecolor{brown}{rgb}{0.59, 0.29, 0.0}
\definecolor{mightnightblue}{RGB}{25,25,112}

\DeclareMathOperator{\Det}{Det}

\def\gsim{\raise0.3ex\hbox{$\;>$\kern-0.75em\raise-1.1ex\hbox{$\sim\;$}}}
\def\lsim{\raise0.3ex\hbox{$\;<$\kern-0.75em\raise-1.1ex\hbox{$\sim\;$}}}

\def\SM{$\mathrm{SU(3)_c \otimes SU(2)_L \otimes U(1)_Y}$ }

\def\SM{$\mathrm{SU(3)_c \otimes SU(2)_L \otimes U(1)_Y}$ }
\def\SU2{$\mathrm{SU(2)_L}$ } 
\def\21{$\mathrm{SU(2)_L \otimes U(1)_Y}$}
\def\d31{\Delta m^2_{31}}
\def\mnu1{m^{\nu}_1}

\makeatletter
\newcommand*{\rom}[1]{\expandafter\@slowromancap\romannumeral #1@}
\makeatother

\newcommand {\red} {\color{red}}

\newcommand{\AddrSoton}{%
  School of Physics and Astronomy, University of Southampton,\\
  Southampton SO17 1BJ, United Kingdom}
\newcommand{\AddrUCI}{%
  Department of Physics and Astronomy, University of California,\\
   Irvine, CA 92697-4575 USA}
\newcommand{\AddrAHEP}{%
  Instituto de Física Corpuscular (IFIC), Universidad de Valencia-CSIC,\\
Paterna (Valencia) E-46980, Spain}

\begin{document}
\bibliographystyle{unsrt}

\title{\boldmath \color{BrickRed} Quark-lepton mass relations from modular flavor symmetry}

\author{Mu-Chun Chen}\email{muchunc@uci.edu}
\affiliation{\AddrUCI}
\author{Stephen F. King}\email{s.f.king@soton.ac.uk}
\affiliation{\AddrSoton}
\author{Omar Medina}\email{Omar.Medina@ific.uv.es}
\affiliation{\AddrAHEP}
\author{Jos\'{e} W. F. Valle}\email{valle@ific.uv.es}
\affiliation{\AddrAHEP}

\begin{abstract}
\vspace{0.5cm}

The so-called Golden Mass Relation provides a testable correlation between charged-lepton and down-type quark masses, that arises in certain flavor models that do not rely on Grand Unification.
Such models typically involve broken family symmetries. In this work, we demonstrate that realistic fermion mass relations can emerge naturally in modular invariant models, without relying on \textit{ad hoc} flavon alignments.
We provide a model-independent derivation of a class of mass relations that are experimentally testable.
These relations are determined by both the Clebsch-Gordan coefficients of the specific finite modular group and the expansion coefficients of its modular forms, thus offering potential probes of modular invariant models.
As a detailed example, we present a set of viable mass relations based on the $\Gamma_4\cong S_4$ symmetry, which have calculable deviations from the usual Golden Mass Relation.

\end{abstract}

\maketitle
\noindent

\section{Introduction}
\label{sec:Introduction}

In the Standard Model (SM), fermion masses and mixings arise from the Yukawa interaction of quarks and leptons with the Higgs field.
Although the fields of the three families have identical SM gauge group quantum numbers, they exhibit largely distinct masses.
Such hierarchical structure of masses across the three families appears rather enigmatic \cite{Xing:2014sja,Feruglio:2015jfa,Xing:2020ijf}.
Moreover, the mixing pattern of quarks and leptons encoded in the CKM and lepton mixing matrices is quite different, and unexplained from first principles in the SM.\par
Understanding the pattern of fermion masses and mixings presents a two-fold puzzle for particle physics.
While some success has been achieved towards predicting fermion mixings through the imposition of family symmetries \cite{King:2013eh,King:2014nza,King:2015aea,King:2017guk,Feruglio:2019ybq},
less progress has been made concerning the formulation of a fully convincing theory of fermion mass hierarchies, though there have been many proposals in this direction \cite{Froggatt:1978nt,Koide:1982ax,Leurer:1992wg,Ibanez:1994ig,Babu:1995hr,Randall:1999ee,Kaplan:2001ga,Chen:2008tc,Buras:2011ph,Weinberg:2020zba}.     \par
The idea of relating quark and lepton masses has a long history. 
Since the $SU(5)$ model proposed by Georgi and Glashow~\cite{Georgi:1974sy}, that places quarks and leptons within a common representation,
it has become usual to expect quark and lepton mass relations to emerge from  gauge unification  \cite{Lazarides:1980nt,Altarelli:1998ns,Ross:2007az,Antusch:2013rxa,Antusch:2015nwi,Antusch:2019avd,Antusch:2022afk,Antusch:2023kli}.\par  
However, despite many efforts that started rather early on~\cite{Georgi:1979df}, no truly definitive theory relating quarks and leptons has ever been devised. 
The  so-called flavor puzzle became more acute after the discovery of neutrino oscillations~\cite{Kajita:2016cak,McDonald:2016ixn}
and the need to account for neutrino masses and mixings as well.\par
Interestingly, viable relations between quark and lepton masses can also emerge in flavor symmetry models,
even in the absence of genuine gauge unification.
This is the case for the so-called approximate \textit{golden} quark-lepton mass relation~\cite{Morisi:2011pt,Bazzocchi:2012ve,King:2013hj,Bonilla:2014xla,Reig:2018ocz}
\begin{equation}
\frac{m_{b}}{\sqrt{m_{s} m_{d}}} \approx \frac{m_{\tau}}{\sqrt{m_{\mu} m_{e}}}~,
\label{eq:GoldenMassRelation}
\end{equation}
that has been obtained both with discrete as well as continuous family symmetry groups.  
This relation has also been obtained in the context of orbifold extensions of the SM~\cite{deAnda:2019jxw,deAnda:2020ssl,deAnda:2020pti,deAnda:2021jzc,deAnda:2022rpw}. 
Given the experimental uncertainty of down-quark mass measurements,
\begin{equation}
\frac{m_d}{m_s} \approx \frac{1}{20}\,,\qquad \frac{m_s}{m_b} \approx \frac{1}{50}\,,\qquad \frac{m_e}{m_\mu} \approx \frac{1}{200}\,, \qquad \text{and}\qquad \frac{m_\mu}{m_\tau} \approx \frac{1}{17}\,,
\label{eq:MassRatios}
\end{equation}
one can readily verify that this mass relation is consistent with experimental data. 
Cleary this is just one relation and, by itself, does not exhaust the complexity of the flavor problem.
However, given its success and simplicity, one may argue that it could constitute part of the ultimate theory of flavor.  
Note also that it is consistent with the Georgi-Jarlskog mass relations~\cite{Georgi:1979df},
\begin{equation}
\frac{m_e}{m_d} \approx \frac{1}{3}\,,\qquad \frac{m_\mu}{m_s} \approx 3\,,\qquad \frac{m_{\tau}}{m_b} \approx 1\,, 
\label{eq:GJ}
\end{equation}
which was predicted to hold at the GUT scale, with quark masses increased by a factor of about 3 at low energies, due to renormalization group (RG) running
(with the largest contribution coming from QCD). 
We emphasize that the combination in Eq.~(\ref{eq:GoldenMassRelation}) (satisfied by the Georgi-Jarlskog relations) is rather stable under renormalization group evolution \cite{Antusch:2013jca,Straub:2018kue}.
As a consequence the Golden mass relation, which holds at the electroweak scale could potentially hold also at high energy scales,
even all the way up to the gauge unification scale $M_{\text{GUT}}\sim 10^{16}$ GeV (See discussion in Subsection \ref{Subsec:Stability}).\par
A common drawback of flavor symmetry model predictions, such as Eq.~(\ref{eq:GoldenMassRelation}), is that they usually rely 
  on \textit{ad hoc} flavor symmetry breaking and vacuum alignment assumptions, for example through \textit{flavons} in the scalar
  sector~\cite{Babu:2002dz,Low:2003dz,Chen:2009um,Barry:2010zk,Altarelli:2010gt,Altarelli:2005yx,Altarelli:2005yp,deAdelhartToorop:2011re,Feruglio:2007uu,Chauhan:2023faf,Chen:2013aya,Altarelli:2010gt,King:2013eh,Feruglio:2009iu}. 
By contrast, in modular invariant models where the flavor symmetry is nonlinearly realized \cite{Feruglio:2017spp} (for a recent review see \cite{Ding:2023htn}),
minimal, realistic, and uniquely defined symmetry breaking patterns can be obtained without the need for flavons.
Moreover, these symmetries could unveil a possible connection between the SM and strings or extra dimensional field theories \cite{Ferrara:1989qb,Ferrara:1989bc,Lauer:1990tm,Cremades:2004wa,Kobayashi:2016ovu,Kobayashi:2018rad,Baur:2019kwi,Baur:2020yjl,Baur:2021bly,Nilles:2023shk,Olguin-Trejo:2018wpw,Ishiguro:2020tmo}.\par 
In this work we point out that modular flavor symmetries can naturally yield viable correlations between the SM fermion masses without invoking flavons nor Grand Unification.  
This illustrates the potential of these symmetries towards the formulation of a successful and experimentally testable flavor theory of fermion masses.  
Mass relations can emerge from the symmetry structure of the vector-valued modular forms that uniquely parametrize the breaking of modular invariance \cite{Feruglio:2017spp,Ding:2020zxw,Liu:2021gwa,Ding:2023htn}. 
In particular, using modular invariance we obtain analytically a generalization of the \textit{golden} quark-lepton mass relation in Eq.~(\ref{eq:GoldenMassRelation}).  \par  

This work is structured as follows: In Section \ref{sec:MassRelations} we present the general method, i.e.
a model-independent derivation of fermion mass relations in modular invariant models that contain few parameters.
In Section \ref{Sec:ModularGamma4Explicit} we present an explicit example for the $\Gamma \cong S_4$ modular group, in which we derive viable mass relations of down-quarks and charged leptons.
We also contrast these to experimental data.
In Section \ref{sec:discussion-outlook-} we argue that, though the chosen example is not intended to be a complete flavor model,
the general method may be useful to build more comprehensive modular symmetry models of flavor, with significantly fewer free parameters than the SM.

\section{Mass relations from modular symmetry} 
\label{sec:MassRelations}

To begin with, let us assume an $\mathcal{N}=1$ supersymmetric theory, whose action is given as
\begin{equation}
\mathcal{S}=\int d^4 x d^2\theta d^2\bar\theta\, \mathcal{K}(\bar{\tau},\bar{\psi},\tau,\psi)+\int d^4 x d^2\theta \mathcal{W}(\tau,\psi)+h.c \,,
\label{eq:SusyAction}
\end{equation}
where $\mathcal{K}$ is the K\"ahler potential and $\mathcal{W}$ is the superpotential. These are functions of the chiral superfields $\tau$ and $\psi$.

\subsection{Modular invariance}
\label{sec:modular-invariance}

Besides the SM gauge symmetry, we require the action ${\mathcal S}$ to be invariant under the modular group $SL(2,\mathbb{Z}) \equiv \Gamma$, 
 so it remains unchanged under the modular transformation \cite{Feruglio:2017spp,Ferrara:1989qb,Ferrara:1989bc,deAdelhartToorop:2011re} 
\begin{equation}
\tau\xrightarrow{\gamma} \frac{a\tau+b}{c\tau+d}
\qquad \text{where} \qquad \gamma=
\begin{pmatrix}
a&b\\
c&d
\end{pmatrix}\, \quad \text{with} \quad ad - cb=1
\label{eq:TauTransformation}
\end{equation}
where the matrix $\gamma$ has integer entries and belongs to the modular group $\Gamma$, generated by the elements 
\begin{equation}
  S = \begin{pmatrix} 0 & 1 \\ -1 & 0 \end{pmatrix} \,, \quad
  T = \begin{pmatrix} 1 & 1 \\ 0 & 1 \end{pmatrix}\,, \quad
  R = \begin{pmatrix} -1 & 0 \\ 0 & -1 \end{pmatrix}\,,
   \label{eq:ModGroupGen}
 \end{equation}
 that obey the relations $S^2 = R$, $(ST)^3 = R^2 = \mathbf{1}$, and $RT = TR$. 
 \par
 The set of matter chiral superfields $\psi$ of the Minimal Supersymmetric Standard Model (MSSM) present in the action in Eq.~(\ref{eq:SusyAction}) transform as weighted representations
 $ \psi \sim \bigoplus_\alpha (\mathbf{r}_\alpha, -k_{\alpha})$ under the action of $\gamma \in \Gamma$.
 \begin{equation}
\psi_\alpha \,\xrightarrow{\gamma} \, 
(c\tau + d)^{-k_\alpha}\, \rho_{\mathbf{r}_\alpha}(\gamma)\, \psi_\alpha\,. 
\label{eq:PsiTransformation}
\end{equation}
Here $\alpha$ labels different irreducible representations $\rho_{\mathbf{r}}(\gamma)$ of $\Gamma_\text{F}$, a finite subgroup  of $\Gamma$ that plays the role of the flavor symmetry.
The corresponding automorphic factor $(c\tau + d)^{-k_\alpha}$ depends on the weight $k_\alpha$.
\par
 The superpotential $\mathcal{W}$, which is assumed to be invariant under a modular transformation, can be written as a power series in the superfields $\psi_{\alpha}$
\begin{equation}
\mathcal{W} = \sum_{n} \left(Y_{\alpha_1\dots\alpha_n} (\tau) \psi_{\alpha_1} \dots \psi_{\alpha_n}\right)_{\mathbf{1}}.
\label{eq:SuperPotExpansion}
\end{equation}
Note that given the transformation properties of $\tau$ and $\psi_\alpha$ under the modular group in Eqs.~(\ref{eq:TauTransformation}) and (\ref{eq:PsiTransformation}) the Yukawa couplings $Y_{\alpha_1\dots\alpha_n} (\tau)$ are requited to be the vector-valued modular forms \cite{Ding:2020zxw,Liu:2021gwa,Liu:2019khw} of the finite modular group $\Gamma_\text{F}$  
\begin{equation}
  Y_{\alpha_1\dots\alpha_n} (\gamma \tau) = (c\tau+d)^{k_{Y_n}} \rho_{Y} (\gamma) Y_{\alpha_1\dots\alpha_n}(\tau),
\label{eq:VectorVModularForms}
\end{equation}
with weight $k_{Y_n}$ and representation $\rho_{Y}(\gamma)$ of $\Gamma_\text{F}$, such that each term of the superpotential in Eq. (\ref{eq:SuperPotExpansion}) is invariant under a modular transformation.\par 
The structure of the K\"ahler potential $\mathcal{K}$, and its covariance under a modular transformation is of great importance once a model is specified~\cite{Chen:2019ewa,Chen:2021prl}.
Nonetheless, since our aim in this paper is not to build a particular modular invariant model but rather to demonstrate a proof of principle that the mass relations can be a result of modular
invariance, we will not include further discussion around it.\par 
The superpotential terms in $\mathcal{W}$ involving the SM fermions can be written compactly as modular invariant fermion bilinears, 
\begin{equation}
    \mathcal{W} \,\supset\, 
    \psi_i \, M(\tau)_{ij} \, \psi^c_j\,, 
    \label{eq:SuperBilinear}
  \end{equation}
  with $i,j=1,\ldots,3$ denoting the three families of the SM. 
Here it is understood that $M({\tau})$ includes the Higgs doublet chiral superfields $\Phi_{u,d}$ of the MSSM.
    Their vacuum expectation values (VEVs) induce the spontaneous breaking of the SM electroweak symmetry, while
    the VEV of the scalar component of the field $\tau$, the modulus, characterizes the breaking of modular invariance. 
    Altogether, the $\Phi_{u,d}$ and $\tau$ VEVs give rise to the mass matrices of the SM fermions. \par 
 \begin{figure}[h!]
    \centering
    \includegraphics[width=0.5\textwidth]{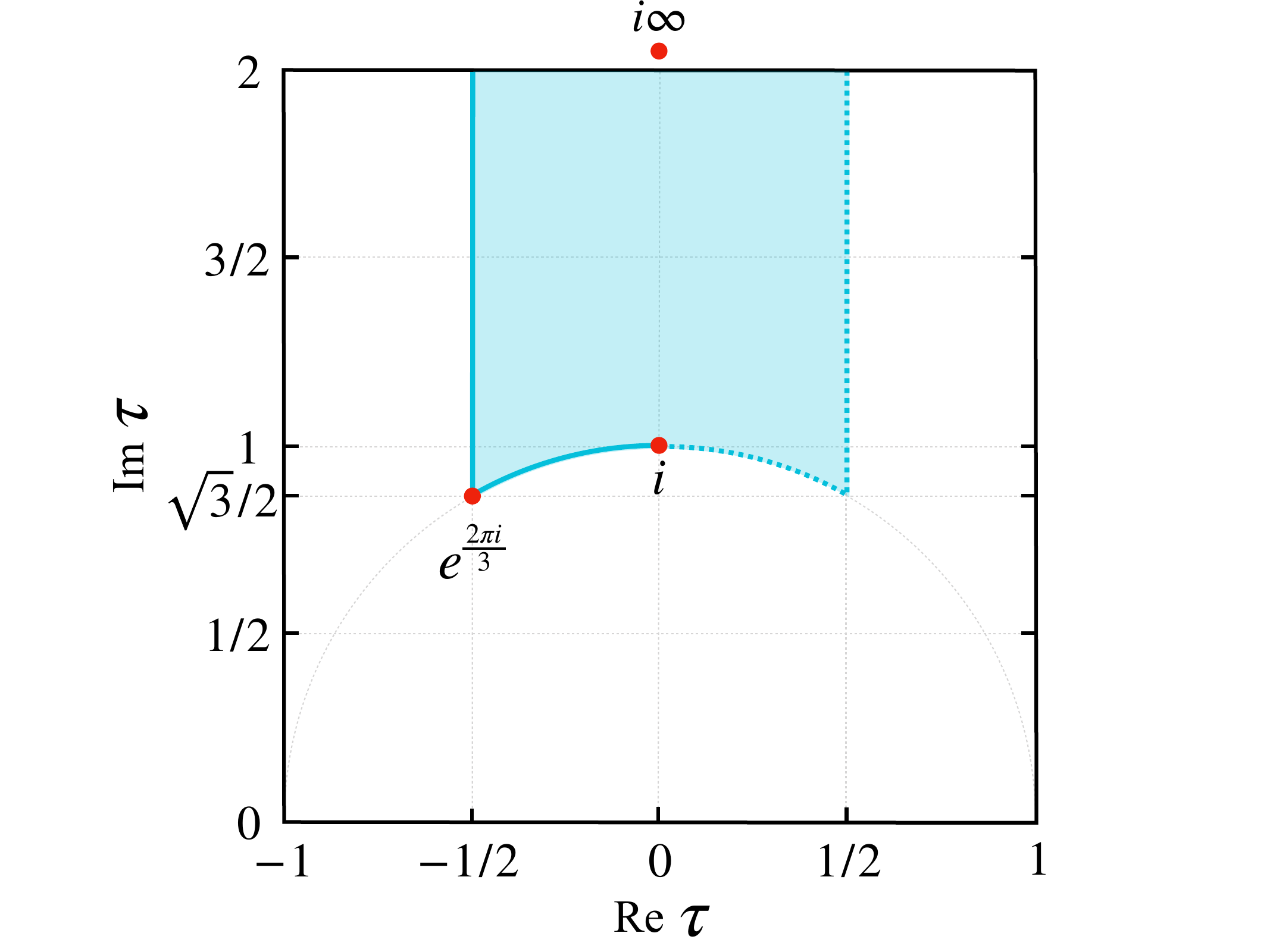} 
    \caption{This figure shows the fundamental domain for the action of the modular group in cyan. We chose to include the solid line border. The three symmetry points, $ \tau_\text{sym}=i\infty,\,i,\,\omega $, are marked in red. An arbitrary value of $ \tau $ in the complex plane can be mapped to a point within this domain via a modular transformation in Eq. (\ref{eq:TauTransformation}).}
    \label{fig:FundamentalDomain}
\end{figure}
  We now turn to the issue of modular symmetry breaking and residual symmetries.
  Note that, as seen from Eqs. (\ref{eq:TauTransformation}) and (\ref{eq:VectorVModularForms}), modular invariance is nonlinearly realized, and any value of $\tau$ will break this symmetry.
  In Fig. \ref{fig:FundamentalDomain} we display the fundamental domain of the modular group action on $\tau$.
Furthermore, there are three special values of the modulus $\tau$ at which the modular group $\Gamma$ breaks down into different preserved residual discrete symmetries \cite{Reyimuaji:2018xvs,Feruglio:2017spp, Novichkov:2018ovf, Ding:2019gof, Novichkov:2021evw}. We generically denote these values as $\tau_{\text{sym}}$ and refer to them as symmetry points.
 These are associated to some unbroken combination of the generators of the modular group in Eq. (\ref{eq:ModGroupGen}). 
As seen from Eqs. (\ref{eq:TauTransformation}) and (\ref{eq:ModGroupGen}) the $R$ generator is unbroken for any value of $\tau$, so that a \(\mathbb{Z}_2^R\) symmetry is always preserved.
Therefore the three symmetry points are 
  \begin{itemize}
\item $\tau_\text{sym} = i \infty$, invariant under $T$, preserving $\mathbb{Z}_N^T \otimes \mathbb{Z}_2^R$.
\item $\tau_\text{sym} = i$, invariant under $S$, preserving $\mathbb{Z}_4^S$, with $\mathbb{Z}_2^R$ as a subgroup.
\item $\tau_\text{sym} = \omega \equiv \exp (2\pi i / 3)$, invariant under $ST$, preserving $\mathbb{Z}_3^{ST} \otimes \mathbb{Z}_2^R$.
\end{itemize}
Notice that for $\tau_\text{sym} = i \infty$ the preserved symmetry $\mathbb{Z}_N^T \otimes \mathbb{Z}_2^R$ is determined by the order $N$ of the $T$ generator ($T^N=\mathbb{I}$)
of the corresponding finite modular group.\par 
When the modulus $\tau$ is near any of the symmetry points $\tau_{\text{sym}}$, certain appealing properties of the modular forms in Eq.~(\ref{eq:VectorVModularForms}) become manifest.
  For instance, they can give rise to hierarchical patterns of fermion masses \cite{Feruglio:2021dte,Novichkov:2021evw,Feruglio:2022koo,Feruglio:2023mii}, a property that is fundamental to this work.
Our main result, outlined below in this section, emerges from these residual symmetries in a model-independent manner. 
Our derivation holds as long as we are close to any of the three symmetry points $\tau_{\text{sym}}= \, i\infty,\,i,\,\, \omega$.
It is convenient to define deviation parameters from each of the three symmetry points 
\begin{equation}
\epsilon_{i\infty}(\tau) = e^{i\frac{2\pi \tau}{N}},\qquad\quad \epsilon_{i}(\tau) = \frac{\tau - i}{\tau + i},\qquad\quad \text{and} \qquad \quad \epsilon_{\omega}(\tau) = \frac{\tau - \omega}{\tau - \omega^2}.
\label{eq:DeviationParameters}
\end{equation}
Note that one can write the modular form multiplets in Eq. (\ref{eq:VectorVModularForms}) in terms of these variables.  
This is useful since the expansion the vector-valued modular forms in terms of $\epsilon_{i\infty}$, $\epsilon_{i}$
or $\epsilon_{\omega}$ is related to their ``charge'' under the respective residual symmetry group, as stressed in \cite{Novichkov:2021evw}.

We now turn to the main result of our proposal, which is a model-independent derivation of how correlations between fermion masses can emerge naturally from  modular flavor symmetries. 
We denote a generic pair of matter superfield multiplets $\psi$ and $\psi^c$ resembling the three SM fermion families as:  
\begin{equation}
\psi = \begin{pmatrix} \psi_1 \\ \psi_2 \\ \psi_3 \end{pmatrix}, \quad \text{and} \quad \psi^c = \begin{pmatrix} \psi^c_1 \\ \psi^c_2 \\ \psi^c_3 \end{pmatrix}.
\label{eq:TripletSuperfields}
\end{equation}
In general, one can write a general fermion mass matrix $M_{\psi}$ as a function of $n$ parameters $a_1,\dots,a_n$ present in the superpotential 
\begin{equation}
\mathcal{W} \, \supset \, \psi M_\psi(a_1,\dots,a_n) \psi^c\hspace{2pt}\hspace{1pt},
\label{eq:MGeneralBilinear}
\end{equation}
so that the three fermion masses are functions of these parameters: 
  \begin{equation}
m_1(a_1,\dots,a_n), \quad m_2(a_1,\dots,a_n), \quad \text{and} \quad m_3(a_1,\dots,a_n).
\label{eq:MassFunctions}
\end{equation}
Given the chiral structure of the MSSM gauge interactions, it is convenient to use the Hermitian and positive semi-definite matrix $H_\psi$ instead of the original mass matrix  
  $M_{\psi}$ in Eq. (\ref{eq:MGeneralBilinear}),
\begin{equation}
H_\psi \equiv M_{\psi} M^{\dagger}_{\psi},
\label{eq:Hpsi}
\end{equation}
this eliminates unphysical mixing parameters associated with the \SU2 singlet superfields $\psi^c$.  
The squared masses of the three fermions are obtained by performing a unitary basis tranformation $U_{\psi}$ over the $\psi$ superfields:   
\begin{equation}
D^2_{\psi} = U^{\dagger}_{\psi} H_{\psi} U_{\psi}, \qquad \text{where} \qquad D^2_{\psi} \equiv \text{Diag}(m^2_1,m^2_2,m^2_3).
\end{equation}
Without loss of generality we assume the following ordering of the masses,
\begin{equation}
m_1 \leq m_2 \leq m_3.
\label{eq:MassOrdering}
\end{equation}
Solving for the individual masses in Eq. (\ref{eq:MassFunctions}) is, in general, a highly non-trivial model-dependent task.   
In contrast, we can use a set of basis invariants of $H_\psi (a_1,\dots,a_n)$ to extract useful information concerning the SM fermion masses in terms of the $a_1,\dots,a_n$ parameters (see for example \cite{Bento:2023owf}), i.e.   
\begin{align}
\text{Tr}[H_\psi] &= m^2_1+ m^2_2 + m^2_3,
\label{eq:TrH}\\
 \frac{1}{2}\left\{ (\text{Tr}[H_\psi])^2-\text{Tr}[H_\psi H_\psi]\right\}&= m^2_1 \, m^2_3+ m^2_2 \, m^2_3 + m^2_1\, m^2_2,
\label{eq:Tr2H}\\
\text{Det}[H_\psi] &= m^2_1 \, m^2_2 \, m^2_3.
\label{eq:DetH}
\end{align}
In these equations, the left-hand side are polynomials in the underlying parameters $a_1,\dots,a_n$,  
with mass-dimension of $2$, $4$, and $6$, respectively. 

\subsection{Mass matrices at the symmetry points} 
\label{subsec:MassMatrixResidual}
Close to the three symmetry points $\tau_\text{sym} = i \infty,\, i, \, \omega $ one can conveniently write $H_{\psi}(a_1,\dots,\epsilon(\tau))$ as a function of the dimensionless parameter $\epsilon(\tau)$
\cite{Feruglio:2021dte,Novichkov:2021evw,Feruglio:2022koo,Feruglio:2023mii}.
In a given model the definition of $\epsilon$ will depend on the corresponding symmetry point $\tau_\text{sym}$ according to the definitions in Eq. (\ref{eq:DeviationParameters}).
We will write the expressions in terms of $\epsilon(\tau)$ instead of $\tau$. \par
In a modular invariant model, the number of independent parameters of the superpotential is reduced, since each term must be invariant under a modular transformation, i.e. 
\begin{equation}
\mathcal{W} \supset \sum_{i} \alpha_i\left(  Y^{(k)}_{\mathbf{r}}(\epsilon)  \psi^{\dagger} \Phi_{u,d} \psi^c \right)_{\mathbf{1}}\,,
\label{eq:SuperPotCon}
\end{equation}
where $\Phi_{u,d}$ denote the two MSSM Higgs doublets, while $Y^{(k)}_{\mathbf{r}}(\epsilon)$ represents the vector-valued modular forms of weight $k$. The index $i$ labels all the modular invariant contractions of $Y^{(k)}_{\mathbf{r}}(\epsilon)$ with the MSSM superfields and $\alpha_i$ are independent numerical coefficients.
To match with our notation in Eq. (\ref{eq:MGeneralBilinear}), we will consider the dimensionful quantities $a_i\equiv \alpha_i v_{u,d}/\sqrt{2}$ as the mass matrix parameters, 
along with $\epsilon$ describing the deviation from one of the three residual symmetry points. 
Note that in top-down constructions the superpotential coefficients $\alpha_i$ are expected to be correlated quantities
\cite{Cremades:2004wa,Baur:2019iai,Nilles:2020tdp,Baur:2020jwc,Almumin:2021fbk,Baur:2022hma,Nilles:2023shk,Baur:2021bly,Kobayashi:2006wq,Abe:2009vi,Kai:2023ivp}, nonetheless in bottom-up constructions they are taken to be independent.\par 
At the symmetry point (in the $\epsilon \to 0$ limit) a discrete residual symmetry is preserved.
For a given finite modular flavor group, and for certain weighted representation assignments for $\psi$ and $\psi^c$, the rank of $H_\psi$ will be reduced in the symmetric limit,
  due to the preservation of the corresponding residual symmetry  \cite{Feruglio:2021dte,Novichkov:2021evw,Feruglio:2022koo,Feruglio:2023mii},
\begin{equation}
\lim_{\epsilon \to 0} \text{Det}[H_{\psi}(a_1,\dots,\epsilon)]= m^2_1\, m^2_2 \,m^2_3 =0.
\label{eq:Det3Lim}
\end{equation}
This condition implies that at least the first family is massless in the symmetric limit.
Thus, the small mass $m_1$ would result from the deviation of the modulus $\tau$ from the residual symmetry point $\tau_\text{sym}$,
providing a symmetry-based explanation for the lightness of the first-family fermion. As detailed in \cite{Novichkov:2021evw}, certain weighted representation assignments can render $m_1$, $m_2$, and even $m_3$ massless at the symmetry point. In our derivation, we focus on the minimal scenario where only $m_1$ is massless in this limit, though the approach is directly applicable to the other cases. 
\subsection{Conditions for mass relations}  
\label{subsec:conditions}

If the number of coefficients $\alpha_i$ is either one ($\alpha_1$) or two ($\alpha_1,\alpha_2$) in Eq. (\ref{eq:SuperPotCon}), the mass matrix resulting from the superpotential will depend on
at most three independent parameters, one of which is dimensionless ($\epsilon$).  
In this case the system in Eqs.~(\ref{eq:TrH})-(\ref{eq:DetH}) is a solvable polynomial system of three variables. \par 
This implies that in a given model, the $H_{\psi}$ matrix will lead to a correlation among the three fermion masses, provided it satisfies the following two conditions:
\begin{enumerate}
\item The superpotential (\ref{eq:SuperPotCon}) must contain (at most) two independent coefficients, $\alpha_1$ and $\alpha_2$,
  so that $H_\psi$ depends only on two dimensionful parameters: $a_1 = (\alpha_1  v_{u,d}/\sqrt{2})$ and $a_2 = (\alpha_2 v_{u,d}/\sqrt{2})$, in addition to $\epsilon$.
  This condition is necessary, otherwise, Eqs. (\ref{eq:TrH})-(\ref{eq:DetH}) form an unconstrained polynomial system. 
\item The three-dimensional matrix $H_\psi$ reduces its rank at the symmetry point, as indicated by Eq. (\ref{eq:Det3Lim}), i.e.
\begin{equation}
\text{ rank} \left[\lim_{\epsilon \to 0}H_{\psi}(a_1,a_2,\epsilon)\right] < \text{rank}[H_{\psi}(a_1,a_2,\epsilon)].
\label{eq:MRankReduction}
\end{equation}
\end{enumerate}
In general, since $H_\psi$ is positive semi-definite, one can write Eq.~(\ref{eq:DetH}) as a power expansion in the dimensionless parameter $|\epsilon|$ around the symmetry point $\tau_\text{sym}$, leading to
\begin{equation}
 \text{Det}[H_{\psi}(a_1,\dots,\epsilon)] = \sum^{\infty}_{m=0} f_m(a_1,\dots, a_{n}) |\epsilon|^{2m}.
\label{eq:DetExpansion}
\end{equation}
If the second condition above is satisfied by $H_{\psi}$ then $f_{0}(a_1,\dots, a_{n})=0$ in the expasion. Therefore, when both conditions hold we can write the last equation as
\begin{equation}
\qquad \qquad \qquad\qquad \text{Det}[H_{\psi}(a_1,a_2,\epsilon)] = m^2_1\, m^2_2\,m^2_3 \equiv f_{\psi} (a_1,a_2) \hspace{2pt} |\epsilon|^{\eta} + \mathcal{O}\left(|\epsilon|^{2\eta}\right)+\dots \hspace{2pt}, \qquad  \text{where}\quad  \eta >0.
\label{eq:DetEpsilon}
\end{equation}
where we defined $f_\psi (a_1,a_2) \equiv f_{\eta}(a_1,a_2)$, which is the coefficient of the leading term in the expansion in Eq. (\ref{eq:DetExpansion}).
From dimensional analysis, we know that $f_\psi$ is a homogeneous order-six polynomial in the parameters $a_1$ and $a_2$.  
However, since $\epsilon$ is dimensionless, the value of $\eta$ can not be inferred solely from dimensional analysis.\par  
Using Eqs. (\ref{eq:MassOrdering}) and (\ref{eq:Det3Lim}), we can write (\ref{eq:TrH}) and (\ref{eq:Tr2H}) in the symmetric limit 
\begin{align}
\lim_{\epsilon \to 0} \text{Tr}[H_\psi] &\equiv h_\psi(a_1, a_2) = m^2_2 + m^2_3,
\label{eq:TrHSym}\\
 \lim_{\epsilon \to 0} \frac{1}{2}\left\{ (\text{Tr}[H_\psi])^2-\text{Tr}[H_\psi H_\psi]\right\} & \equiv g_\psi(a_1,a_2) =  m^2_2\, m^2_3.
\label{eq:Tr2HSym}
\end{align} 
The two homogeneous polynomials $h_\psi(a_1,a_2)$ and $g_\psi(a_1,a_2)$ are of order $2$ and $4$ respectively. This is a two-equation system of two variables $a_1$ and $a_2$.
At the exact symmetry point these polynomials give solutions of the form  
\begin{align}
\tilde{a}_{1}(m_2, m_3)\, , \quad \text{and} \quad  \tilde{a}_{2}(m_2, m_3)\,,
\label{eq:Sola1}
\end{align}
which relate the model parameters $a_1$, $a_2$ with the fermion masses $m_2$ and $m_3$. See Appendix \ref{Appendix:Polynomials} for further discussion about the polynomial system solutions.\par  
The solutions in the last equation, that hold at the symmetry point ($\epsilon \to 0$), together with Eq. (\ref{eq:DetEpsilon}) lead us to define the polynomial 
\begin{equation}
f(m_2,m_3) \equiv f_{\psi}(\tilde{a}_{1}(m_2, m_3),\tilde{a}_{2}(m_2, m_3)),
\label{eq:fpolyMasses}
\end{equation}
Close to the symmetry point we can approximate Eq. (\ref{eq:DetEpsilon}) to its leading term in $ \lvert \epsilon \rvert \ll1$. This yields an expression relating the three fermion masses to $ \lvert \epsilon \rvert$ 
\begin{equation}
\frac{m^2_1\, m^2_2\,m^2_3}{ f (m_2,m_3) } \approx |\epsilon|^{\eta} \,.
\label{eq:MassesAndEpsilon}
\end{equation}
This is our central result, it
allows us to identify potentially viable correlations between the masses of the SM fermions in modular invariant models. 
In a specific model, the polynomial $f(m_2,m_3)$ is determined by both the Clebsch-Gordan coefficients of the finite modular group $\Gamma_\text{F}$ and the vector-valued modular forms in the superpotential terms of Eq. (\ref{eq:SuperPotCon}).
It is important to note that Eq. (\ref{eq:MassesAndEpsilon}) serves as a valid approximation near the symmetry points $\tau_\text{sym} = i \infty,\, i, \, \omega $; otherwise, additional terms of the expansion in Eq. (\ref{eq:DetEpsilon}) must be considered.
Notice that Eq. (\ref{eq:MassesAndEpsilon}) is fully independent of the phase of $\epsilon(\tau)$.

\section{Example of viable mass relations} 
\label{Sec:ModularGamma4Explicit}

To illustrate the general derivation in the previous section and emphasize key points, we now turn to a specific example. 
Let's consider $\Gamma_4 \cong S_4$ as the finite modular group. There are several flavor models based on this group or its double cover (see e.g. \cite{Penedo:2018nmg,Liu:2020akv,Qu:2021jdy,Abe:2023qmr,deMedeirosVarzielas:2023crv,Abe:2023ilq,Novichkov:2020eep,Criado:2019tzk,Novichkov:2018ovf}).
Detailed information about the group properties, the chosen basis for the representations, and the expansion of the relevant modular forms are given in Appendix \ref{Appendix:Gamma4Modular}. \par

The $\Gamma_4$ weighted-representation assignments of the down-sector superfields that result in a quark-lepton mass relation are given in Table~\ref{tab:S4IrrepExample}.
For simplicity we omit transformation properties under the \SM gauge symmetry, as they are the standard ones of the MSSM \cite{Martin:1997ns}. 
As mentioned in the previous section, the first necessary condition for predicting a fermion mass relation is related to the number of invariant contractions in the superpotential (see Eq. (\ref{eq:SuperPotCon})).
In the case of $\Gamma_4$ at weight $2$, there are five modular forms arranged in two multiplets: 
\begin{equation} 
Y^{(2)}_{\mathbf{2}}(\tau)=\begin{pmatrix}Y_1(\tau)\\ Y_2(\tau)\end{pmatrix},\hspace{10pt} Y^{(2)}_{\mathbf{3}}(\tau)=\begin{pmatrix}Y_3(\tau)\\ Y_4(\tau)\\  Y_5(\tau)\end{pmatrix}.
\label{eq:S4ModularFormsWeight2}
\end{equation}
Hence there are only two independent contractions of the superfields with the modular forms, 
\begin{table}[h]
\centering
\begin{tabular}{|c|c|c|c|c|c|c|c|}
\hline
\textbf{}      & $Q$          & $D^c$        & $L$          & $E^c$        & $\Phi_d$        & $Y^{(2)}_{\mathbf{2}}$ & $Y^{(2)}_{\mathbf{3}}$ \\ \hline
\textbf{$\Gamma_4$} & $\mathbf{3}$ & $\mathbf{3^{\prime}}$ & $\mathbf{3}$ & $\mathbf{3^{\prime}}$ & $\mathbf{1}$ & $\mathbf{2}$           & $\mathbf{3}$           \\ \hline
\textbf{$k$}   & $-1$         & $-1$         & $-2$         & $0$         & $0$         & $2$                    & $2$                    \\ \hline
\end{tabular}
\caption{This table contains the $\Gamma_4$ weighted-representation assignments that render a quark-lepton mass relation. MSSM gauge symmetry transformations are omitted in this table.}
\label{tab:S4IrrepExample}
\end{table}
\begin{equation}
\mathbf{3} \otimes \mathbf{3^{\prime}} \otimes Y^{(2)}_{\mathbf{2}} \Longrightarrow \mathbf{1} \qquad \text{and} \qquad \mathbf{3} \otimes \mathbf{3^{\prime}} \otimes Y^{(2)}_{\mathbf{3}} \Longrightarrow \mathbf{1} \hspace{3pt}.
\label{eq:S4ModelProd}
\end{equation}
Hence, the relevant superpotential terms in Eq. (\ref{eq:SuperPotCon}) for down-quarks and charged-leptons have the following structure:
\begin{align}
\mathcal{W}^{\Gamma_4}_{\Phi_d} &\supset \hspace{3pt} \alpha^{d}_{1} \left(Q \Phi_d D^c Y^{(2)}_{\mathbf{2}}(\tau) \right)_{\mathbf{1}} + \hspace{3pt}  \alpha^{d}_{2} \left(Q \Phi_d D^c Y^{(2)}_{\mathbf{3}}(\tau) \right)_{\mathbf{1}} \nonumber\\
 &\hspace{3pt}+\alpha^{e}_{1} \left(L \Phi_d E^c Y^{(2)}_{\mathbf{2}}(\tau) \right)_{\mathbf{1}} + \hspace{3pt} \alpha^{e}_{2} \left(L \Phi_d E^c Y^{(2)}_{\mathbf{3}}(\tau) \right)_{\mathbf{1}}.
\label{eq:YukawaS4}
\end{align}
There are four coefficients in Eq.~(\ref{eq:YukawaS4}): $\alpha^{d}_{1}$, $\alpha^{d}_{2}$, $\alpha^{e}_{1}$, and $\alpha^{e}_{2}$. In our bottom-up setup, these are independent parameters. 
Although they are potentially complex, we assume the preservation of CP symmetry so these coefficients are real \cite{Novichkov:2019sqv}. 
We define the following set of dimensionful parameters 
\begin{equation}
a^d_1 \equiv \frac{\alpha^d_1 v_d}{\sqrt{2}}, \quad a^d_2 \equiv \frac{\alpha^d_2 v_d}{\sqrt{2}}, \quad a^e_1 \equiv \frac{\alpha^e_1 v_d}{\sqrt{2}}, \quad \text{and} \quad a^e_2 \equiv \frac{\alpha^e_2 v_d}{\sqrt{2}}.
\label{eq:ParameterMassDimension}
\end{equation}
in terms of the standard MSSM VEVs $\langle \Phi_{u,d} \rangle \equiv v_{u,d}/\sqrt{2}$\hspace{2pt}.

The mass matrices for down-type quarks and charged-leptons, denoted as $M_d$ and $M_e$ respectively, are given as:
\begin{equation}
M_{f} = 
\begin{pmatrix}
- a^{f}_{1} Y_{2}  & -a^{f}_{2} Y_{4} & a^{f}_{2} Y_{5} \\
-a^{f}_{2} Y_{4} & \frac{1}{\sqrt{2}}\left( \sqrt{3} a^{f}_{1} Y_{1} - 2 a^{f}_{2} Y_{3}\right) & \frac{1}{2} a^{f}_{1} Y_{2} \\
a^{f}_{2} Y_{5}  & \frac{1}{2} a^{f}_{1} Y_{2} &  \frac{1}{\sqrt{2}}\left( \sqrt{3} a^{f}_{1} Y_{1} + 2 a^{f}_{2} Y_{3}\right) \\
\end{pmatrix}, \qquad \text{for} \quad f=\hspace{1pt}d,\hspace{1pt}e\hspace{1pt}.
\label{eq:MdAndMeMatrix}
\end{equation}
For the sake of compactness, we have omitted the explicit $\tau$ dependence of the modular forms $Y_{i}(\tau)$, with $i=1,..,5$. 
Notice that $M_d$ and $M_e$, as described in Eq. (\ref{eq:MdAndMeMatrix}), have a common structure
arising from their shared covariance properties under the gauge as well as the $\Gamma_4$ symmetries. Both mass matrices satisfy our first condition in Section~\ref{subsec:conditions}.\par
As prescribed in the previous section we define the Hermitian matrices
\begin{equation}
H_d \equiv M_d M^{\dagger}_d \hspace{2pt}, \qquad \text{and} \qquad H_e \equiv M_e M^{\dagger}_e \hspace{2pt}.
\label{HdAndHeDef}
\end{equation}
For $\tau$ close to the $T$-symmetric point $\tau_{T} \equiv i \infty$ these matrices fulfill the second condition outlined in Eq. (\ref{eq:MRankReduction}). Consequently, they will lead to mass correlations that we will now discuss. Following Eq. (\ref{eq:DeviationParameters}), throughout this section we define the $\epsilon(\tau)$ parameter: 
\begin{equation}
\epsilon(\tau) \equiv q ^{\frac{1}{4}}=  e^\frac{2\pi i \tau}{4}, \qquad \text{thus} \qquad  \lim_{\tau \to i \infty} \epsilon(\tau) =0.
\label{eq:epsilon}
\end{equation}
In this limit the superpotential preserves a residual $\mathbb{Z}^{T}_4$ symmetry, generated by $T$ in Table \ref{tab:Gamma4GenRep} of Appendix \ref{Appendix:Gamma4Modular}.\par 
Given that our general derivation  applies near the residual symmetry points, we can truncate the expansion of the modular forms given in Appendix \ref{Appendix:Gamma4Modular}
to their leading order in $\epsilon$:
\begin{equation}
Y^{(2)}_{\mathbf{2}}(\epsilon)\approx \begin{pmatrix}1 + {\red 24} \epsilon^4\\ {\red -8 \sqrt{3}} \epsilon^2\end{pmatrix},\hspace{10pt} Y^{(2)}_{\mathbf{3}}(\epsilon) \approx \begin{pmatrix}1 {\red- 8}\epsilon^4 \\ {\red-4 \sqrt{2}} \epsilon\\  {\red-16 \sqrt{2}} \epsilon^3 \end{pmatrix}.
\label{eq:S4Weight2epsilon}        
\end{equation}
It is manifest that the $T$-symmetric limit $\epsilon \to 0$ corresponds to the following alignment of the vector-valued modular forms  
\begin{equation}
\lim_{\epsilon \to 0} Y^{(2)}_{\mathbf{2}}(\epsilon)=\begin{pmatrix}1\\ 0\end{pmatrix},
\hspace{30pt} \lim_{\epsilon \to 0} Y^{(2)}_{\mathbf{3}}(\epsilon)=\begin{pmatrix}1\\ 0\\  0\end{pmatrix}.
\label{eq:YinLimit}
\end{equation}
At this symmetry point we obtain the following mass spectrum for charged-leptons and down-quarks 
\begin{equation}
\lim_{\epsilon \to 0}  \begin{pmatrix} m_{\tau} \\ m_{\mu} \\ m_e \end{pmatrix} = \begin{pmatrix} m_{\tau}\left(a^e_1,a^e_2\right)\\ m_{\mu}\left(a^e_1,a^e_2\right) \\ 0 \end{pmatrix}, \qquad \quad \lim_{\epsilon \to 0}  \begin{pmatrix} m_b \\ m_s \\ m_d \end{pmatrix} = \begin{pmatrix} m_b\left(a^d_1,a^d_2\right)\\ m_s\left(a^d_1,a^d_2\right) \\ 0 \end{pmatrix},
\label{eq:FixedPointMases}
\end{equation}
meaning that in the limit $\tau \to i \infty$ the down-quark and the electron become massless, hence their mass is only generated through a deviation from this symmetry point. 
In this limit, the masses of the the second and third family fermions are functions of the parameters defined in Eq. (\ref{eq:ParameterMassDimension}).\par
Following the general derivation in Section \ref{sec:MassRelations} we determine the polynomials defined in Eqs. (\ref{eq:TrHSym}), and (\ref{eq:Tr2HSym}) for this example: 
\begin{equation}
h_d(a^d_1, a^d_2) = \frac{3}{2}{a^d_1}^2 + 2 {a^d_2}^2 = m^2_s + m^2_b, \qquad g_d(a^d_1, a^d_2) = \frac{1}{16} \left( 3 {a^d_1}^2 -4 {a^d_2}^2\right)^2= m^2_s\, m^2_b, 
\label{eq:HdPoly}
\end{equation}
and analogously for charged-leptons $h_e(a^e_1, a^e_2)$ and $g_e(a^e_1, a^e_2)$ which in this case have the equivalent structure.\par  
We can obtain solutions for these two polynomial systems as described in Eq. (\ref{eq:Sola1}).
For a detailed list of all the different solutions to this polynomial system please refer to Appendix \ref{Appendix:Polynomials}.
Notably, the following set of solutions (obtained for $\epsilon \to 0$) produces viable mass relations: 
\begin{align}
\tilde{a}^d_{1,\pm}(m_s,m_b) &= \frac{m_b\pm m_s}{\sqrt{3}}, \quad \tilde{a}^d_{2,\pm}(m_s,m_b) = \frac{m_b\mp m_s}{2},
\label{eq:ViableSolsDown}\\
\tilde{a}^e_{1,\pm}(m_\mu,m_\tau) &= \frac{m_\tau\pm m_\mu}{\sqrt{3}}, \quad \tilde{a}^e_{2,\pm}(m_\mu,m_\tau) = \frac{m_\tau \mp m_\mu}{2}.
\label{eq:ViableSolsLep}
\end{align}
When the modulus value $\tau$ departs from a given symmetry point, in this case $\tau_{T}= i \infty$, there is  a common scaling dependence on the parameter $\epsilon(\tau)$ that generates both $m_d$ and $m_e$.
Referring to Eq. (\ref{eq:DetEpsilon}) we can express the equations for the determinants of $H_e$ and $H_d$ as an expansion in the $\epsilon$ parameter: 
\begin{align}
\Det\left[ H_d\right]  &= m^2_b \hspace{1pt} m^2_s\hspace{1pt} m^2_d \approx
\frac{4}{27} \left(3 a^d_1 + 2 \sqrt{3} a^d_2\right)^4\left(3 a^d_1 - 4 \sqrt{3} a^d_2\right)^2 |\epsilon|^4 +  \mathcal{O}\left(|\epsilon|^8\right)
\hspace{1pt}.\label{eq:EqDetDown} \\
\Det\left[ H_e\right]  &= m^2_\tau \hspace{1pt} m^2_\mu\hspace{1pt} m^2_e \approx
\frac{4}{27} \left(3 a^e_1 + 2 \sqrt{3} a^e_2\right)^4\left(3 a^e_1 - 4 \sqrt{3} a^e_2\right)^2 |\epsilon|^4 +  \mathcal{O}\left(|\epsilon|^8\right)
\hspace{1pt}.\label{eq:EqDetLep} 
\end{align}
Therefore close to the symmetry point ($|\epsilon| \ll 1$) we can plug in Eqs. (\ref{eq:ViableSolsDown}) and (\ref{eq:ViableSolsLep}) into Eqs. (\ref{eq:EqDetDown}), and (\ref{eq:EqDetLep})  to obtain 
\begin{equation}
m^2_b \hspace{1pt} m^2_s\hspace{1pt} m^2_d \approx f_{\pm}(m_b,m_s)\hspace{2pt} \lvert\epsilon\lvert^4 \hspace{2pt}, \qquad m^2_{\tau} \hspace{1pt} m^2_{\mu}\hspace{1pt} m^2_e \approx f_{\pm}(m_{\tau},m_{\mu})\hspace{2pt} \lvert\epsilon\lvert^4 \hspace{2pt},
\label{eq:FunctionRel}
\end{equation}
where $f_{\pm}(m_3,m_2)$ are  two polynomials of order 6 that depend on masses of the third and second generation fermions. 
The viable polynomial obtained $f_{\pm}(m_3,m_2)$ is given by 
\begin{equation}
f_{\pm}(m_3,m_2)={\red 64} m^4_3\left( m_3 {\mp \red3} m_2 \right)^2.
\label{eq:Gamma4PolyLeading}
\end{equation}

We stress that the coefficients (marked in red in the last equation) in the polynomials are determined not only by the scaling properties of the modular forms in Eq. (\ref{eq:S4Weight2epsilon}) with respect to $ \lvert \epsilon \rvert$
(away from a symmetry point), but they also involve Clebsch-Gordan coefficients of the $S_4$ group and the coefficients of the Fourier expansion of the modular forms (also marked in red in Eq. (\ref{eq:S4Weight2epsilon})).
The expansion coefficients have been proven to contain mathematical information (see e.g.~\cite{bruinier20081}) whose significance in models with modular flavor symmetry remains somewhat underexplored,
as the focus so far has primarily centered on the $\epsilon$ scaling properties \cite{Feruglio:2021dte,Novichkov:2021evw,Feruglio:2022koo,Feruglio:2023mii,Petcov:2023fwh}.\par
\begin{figure}[b!]
    \centering
    \includegraphics[width=0.7\textwidth]{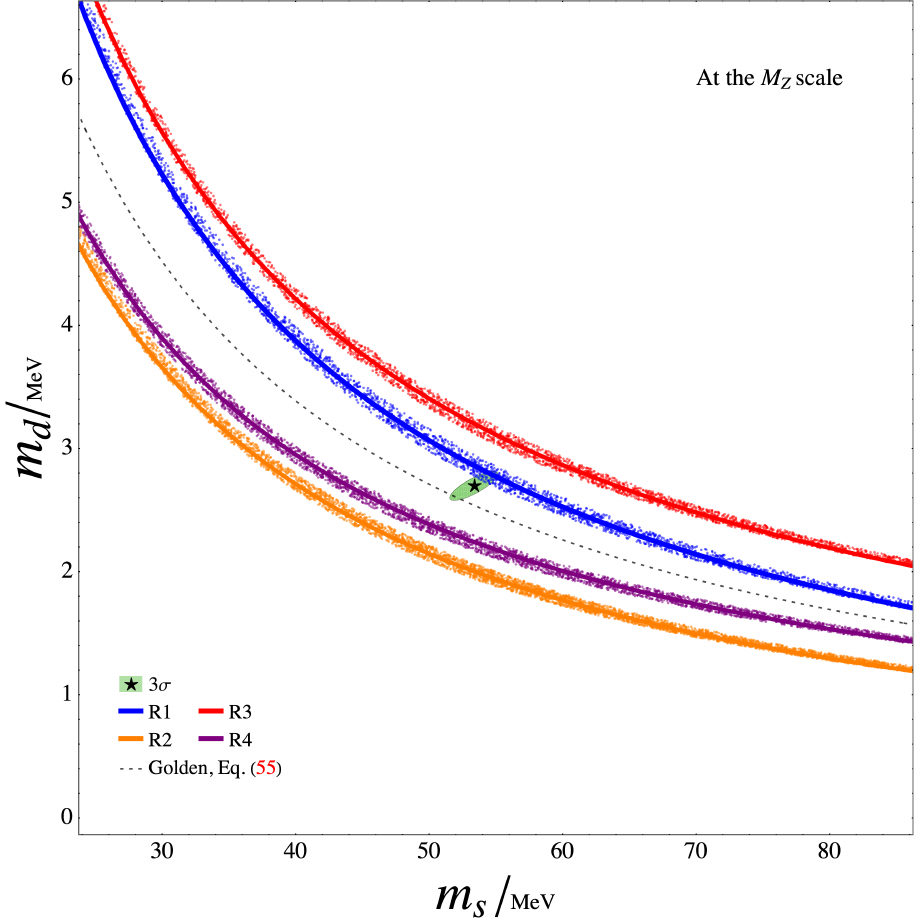} 
    \caption{This plot displays the results of our parameter scan $\{a^{d,e}_{1,2}, \lvert \epsilon \rvert \} \Longrightarrow \{m_{d,s,b,e,\mu,\tau} \}$ in the $m_d-m_s$ plane.
      Each point lies inside the experimental $3$-$\sigma$ range for $m_b$, $m_e$, $m_\mu$, and $m_\tau$.
      The allowed 3-$\sigma$ region for $m_d$ and $m_s$ is shown in light-green \cite{Straub:2018kue}.
      We include the curves of the mass relations R1 to R4 from Eqs. (\ref{eq:MassRelMM}) - (\ref{eq:MassRelPP}) as solid lines. %
      There can be sizable differences between these mass relations.
      For comparison we also plot (dashed) the \textit{golden} quark-lepton mass relation in Eq.~(\ref{eq:MassRel2}) obtained in "non-modular" flavor models~\cite{Morisi:2011pt,Bazzocchi:2012ve,King:2013hj,Bonilla:2014xla,Reig:2018ocz}.}
    \label{fig:MassRelation-Outer}
\end{figure}

\begin{figure}[h!]
    \centering
    \includegraphics[width=0.7\textwidth]{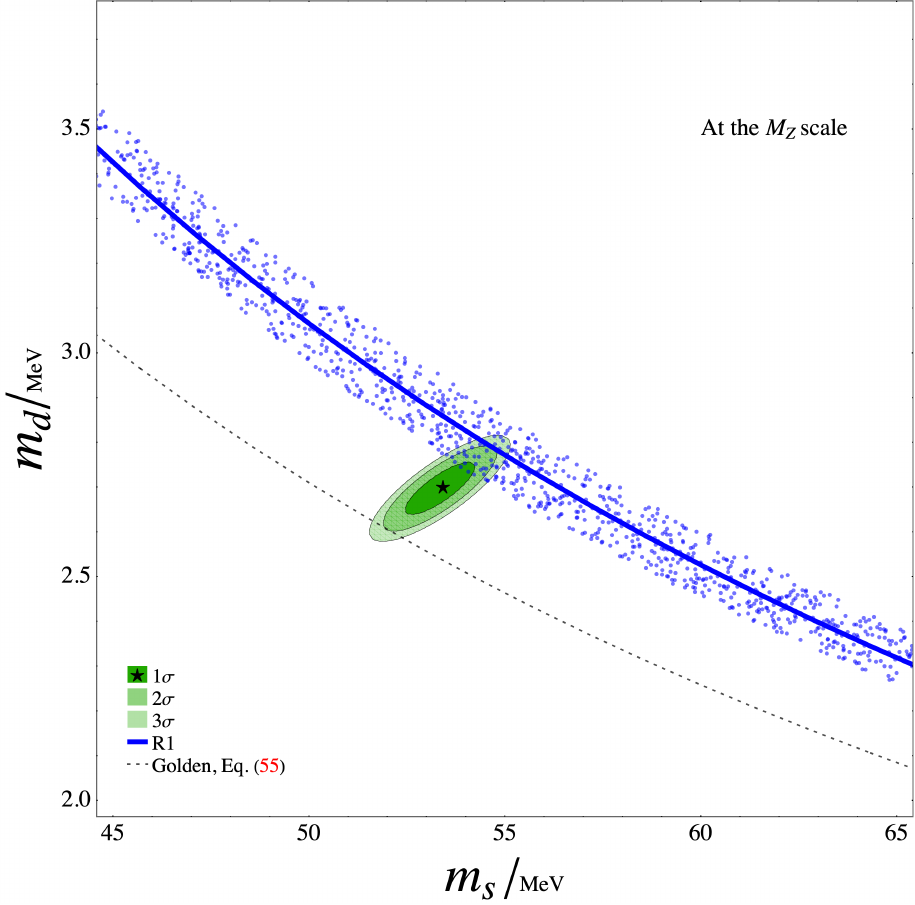} 
    \caption{This is a close-up of Fig. \ref{fig:MassRelation-Outer} where we only display the results for the R1 mass relation in Eq. (\ref{eq:MassRelMM})
      which is the experimentally most favored one. The allowed 1-  2- and 3-$\sigma$ regions for $m_d$ and $m_s$ are shown in different shades of green \cite{Straub:2018kue}. The dashed curve in the plot shows the \textit{golden} quark-lepton mass prediction in Eq.~(\ref{eq:MassRel2}) obtained in "non-modular" flavor
      models~\cite{Morisi:2011pt,Bazzocchi:2012ve,King:2013hj,Bonilla:2014xla,Reig:2018ocz}. An improved determination of $m_s$ and $m_d$ could probe our mass relations.  }
    \label{fig:MassRelation-Inner}
\end{figure}

  
Looking at Eq. (\ref{eq:FunctionRel}) it is straightforward to obtain the predicted mass relations for this example
\begin{equation}
\frac{m^2_b \hspace{1pt} m^2_s\hspace{1pt} m^2_d}{f_{\pm}(m_b,m_s)} \approx \frac{m^2_{\tau} \hspace{1pt} m^2_{\mu}\hspace{1pt} m^2_e}{f_{\pm}(m_{\tau},m_{\mu})}
\label{eq:S4GeneralMassRelation}
\end{equation}
which can be simplified to obtain  
\begin{align}
\frac{ m_s\hspace{1pt} m_d}{ m_b\left( m_b \pm 3 m_s \right)} \approx \frac{ m_\mu \hspace{1pt} m_e}{ m_\tau \left( m_\tau \pm 3 m_\mu \right)} .\label{eq:MassRel}
\end{align}
Different possible signs in the denominators correspond to different solutions in Eqs. (\ref{eq:ViableSolsDown}) and (\ref{eq:ViableSolsLep}). 
There are four distinct mass relations emerging from Eq. (\ref{eq:MassRel}), which we label as follows: 
\begin{align}
\text{R1:}\quad \frac{ m_s\hspace{1pt} m_d}{ m_b\left( m_b - 3 m_s \right)} \approx \frac{ m_\mu \hspace{1pt} m_e}{ m_\tau \left( m_\tau - 3 m_\mu \right)} .\label{eq:MassRelMM}\\
\text{R2:}\quad \frac{ m_s\hspace{1pt} m_d}{ m_b\left( m_b - 3 m_s \right)} \approx \frac{ m_\mu \hspace{1pt} m_e}{ m_\tau \left( m_\tau + 3 m_\mu \right)} .\label{eq:MassRelMP}\\
\text{R3:}\quad \frac{ m_s\hspace{1pt} m_d}{ m_b\left( m_b + 3 m_s \right)} \approx \frac{ m_\mu \hspace{1pt} m_e}{ m_\tau \left( m_\tau - 3 m_\mu \right)} . \label{eq:MassRelPM}\\
\text{R4:}\quad \frac{ m_s\hspace{1pt} m_d}{ m_b\left( m_b + 3 m_s \right)} \approx \frac{ m_\mu \hspace{1pt} m_e}{ m_\tau \left( m_\tau + 3 m_\mu \right)} . \label{eq:MassRelPP}
\end{align}
To scrutinize the viability of these  mass relations we performed a scan over the following set of parameters determining the masses of down quarks and charged leptons 
\begin{equation}
\{a^{d,e}_{1,2}, \lvert \epsilon \rvert \} \Longrightarrow \{m_{d,s,b,e,\mu,\tau} \}.
\label{eq:ParametersQuarkLepton}
\end{equation}
Notice that the light down-quark masses $m_d$ and $m_s$ are determined through lattice QCD computations, and exhibit the largest uncertainties amongst all fermion masses in Eqs. (\ref{eq:MassRelMM})-(\ref{eq:MassRelPP}).
In Figs.~\ref{fig:MassRelation-Outer} and  \ref{fig:MassRelation-Inner} we display the results of our parameter scan, each point plotted falls within the experimental $3$-$\sigma$ region for $m_b$, $m_e$, $m_\mu$, and $m_\tau$. 
In Fig. \ref{fig:MassRelation-Outer} we display the curves corresponding to all four mass relations R1 to R4 derived above with solid lines of different colors,  illustrating how the mass relations differing only on signs in the denominators can be substantially different.
In Fig. \ref{fig:MassRelation-Inner} we show a close-up plot of R1 which is the most favored experimentally.
These curves use the experimental  central values for $m_b$, $m_e$, $m_\mu$, and $m_\tau$, allowing us to compare the analytical mass relations (curves) with the numerical results from the parameter scan (points).
The scatter plot  and the plotted curves show excellent agreement, indicating that the approximations we used to derive the mass relations are accurate enough.\par 

\subsection{Zooming in the golden quark-lepton mass relation} 
\label{sec:zooming-golden-quark}

We will now take a closer look at the derived mass relations in  Eqs. (\ref{eq:MassRelMM})-(\ref{eq:MassRelPP}) in comparison to the golden quark-lepton mass relation obtained from traditional flavor symmetries~\cite{Morisi:2011pt,Bazzocchi:2012ve,King:2013hj,Bonilla:2014xla,Reig:2018ocz}. %
In the usual derivation of the latter, using flavons for example, there are small corrections which are usually neglected, i.e.
\begin{align}
  \frac{ m_b\hspace{1pt}  m_s\hspace{1pt} m_d}{ m_b^3 +\mathcal{O}(m_s^3)} \approx \frac{m_\tau \hspace{1pt}  m_\mu \hspace{1pt} m_e}{ m^3_\tau +\mathcal{O}(m^3_{\mu}) } .
  \label{eq:MassRel2}
\end{align}
These corrections in the denominators of Eq.~(\ref{eq:MassRel2}) are very small.
In contrast, the corrections in Eq. (\ref{eq:MassRel}) arising from Clebsch-Gordan coefficients of the $\Gamma_4$ symmetry and those of the modular forms expansion around the symmetry point, are typically much larger.
They can yield substantial differences, as shown in Figs.~\ref{fig:MassRelation-Outer} and \ref{fig:MassRelation-Inner},
where we also display the prediction from Eq.~(\ref{eq:MassRel2}).  \par

Indeed, the mass relations in Eqs. (\ref{eq:MassRelMM})-(\ref{eq:MassRelPP}) have an interesting property. 
They involve polynomials $f_{\pm}(m_3,m_2)$, Eq. (\ref{eq:Gamma4PolyLeading}), which include a term proportional to $m^6_3$.  
Given the strong hierarchy between the masses of fermions in the second and third families ($m_3 \gg m_2$), these polynomials $f_{\pm}(m_3,m_2)$
can be  approximated using their leading term: 
\begin{equation}
\qquad  \frac{m_2}{m_3} \ll 1 \quad \Longrightarrow  \quad f_{\pm}(m_3,m_2)\approx 64 \hspace{1.5pt}  m^6_3 .
\label{eq:S4PolyApprox}
\end{equation}
This illustrates that there is a class of polynomials $f(m_3,m_2)$, defined in Eq. (\ref{eq:fpolyMasses}),
which have a leading term proportional to $m^{6}_3$. And when used to relate charged-leptons and down-type quarks, this class of polynomials will approximate the \textit{golden} quark-lepton mass relation in Eq.~(\ref{eq:MassRel2}) for largely hierarchical fermion masses $m_1 \ll m_2 \ll m_3$.
In fact, by using this insight, one can infer that the polynomials in Eq. (\ref{eq:Gamma4PolyLeading}) will be roughly consistent with experimental data
even without performing a quantitative parameter scan.\par
What is important to note is that the mass relations in Eqs. (\ref{eq:MassRelMM})-(\ref{eq:MassRelPP}) are indeed valid for the values of quark and lepton masses
(up to theoretical or experimental uncertainty).
Our general derivation in Section \ref{sec:MassRelations} and the above example demonstrate that a class of  viable mass predictions for down-quarks and charged-leptons can arise from modular invariant models near the symmetry points.
Although the obtained mass relation can be similar to the non-modular \textit{golden} mass relation, these predictions can yield testable deviations from it.
As a result, improving determinations of light-quark masses could help constrain modular invariant flavor models,
where the mass hierarchies of quarks and leptons emerge from deviations from residual symmetry points. 

\subsection{Stability under renormalization group evolution} 
\label{Subsec:Stability}

Mass relations are generally very sensitive to renormalization group (RG) running. 
For instance, the Georgi-Jarlskog mass relations in Eq. (\ref{eq:GJ}), derived at the GUT scale~\cite{Georgi:1979df}, undergo significant evolution to fit experimental
quark and lepton masses at lower scales, such as $ M_Z $.
In contrast, the mass relations in Eqs. (\ref{eq:MassRelMM})-(\ref{eq:MassRelPP}), derived from Eq. (\ref{eq:MassesAndEpsilon}),
involve ratios of down-quark to charged-lepton masses. Thus, they are expected to be fairly stable under RG running. 
This follows from the fact that the dominant contribution to the Yukawa couplings' \(\beta\)-functions comes from gauge couplings, which effectively cancel in the mass ratios. However, for
the bottom quark mass $m_b$ there is a contribution involving the top quark Yukawa coupling $y_t$, making it necessary to analyze more closely the stability of our fermion mass relations.  \par
To illustrate this point with an example we define the quantity $ F_{\text{R1}}(\mu_E)$
and rewrite the mass relation R1 in Eq. (\ref{eq:MassRelMM}) with the down-quark and lepton masses given as functions of the scale $\mu_\text{E}$,
\begin{equation}
 F_{\text{R1}}(\mu_E) \equiv \frac{m_s  m_d m_\tau(m_\tau -3 m_\mu)}{m_\mu m_e m_b(m_b -3 m_s)} \approx 1.
\label{eq:MassRelSqrd}
\end{equation}
For definiteness we perform the RG running of the mass relation R1 in the constrained MSSM (CMSSM) \cite{Hall:1983iz,Ghosh:2012dh,Gupta:2020whs},
using the Mathematica Package REAP \cite{Antusch:2005gp} with the extension SusyTC \cite{Antusch:2015nwi} so as to account for SUSY threshold corrections
\cite{Hempfling:1993kv,Blazek:1995nv,Antusch:2013jca}.
\begin{figure}[h!]
    \centering
    \includegraphics[width=0.6\textwidth]{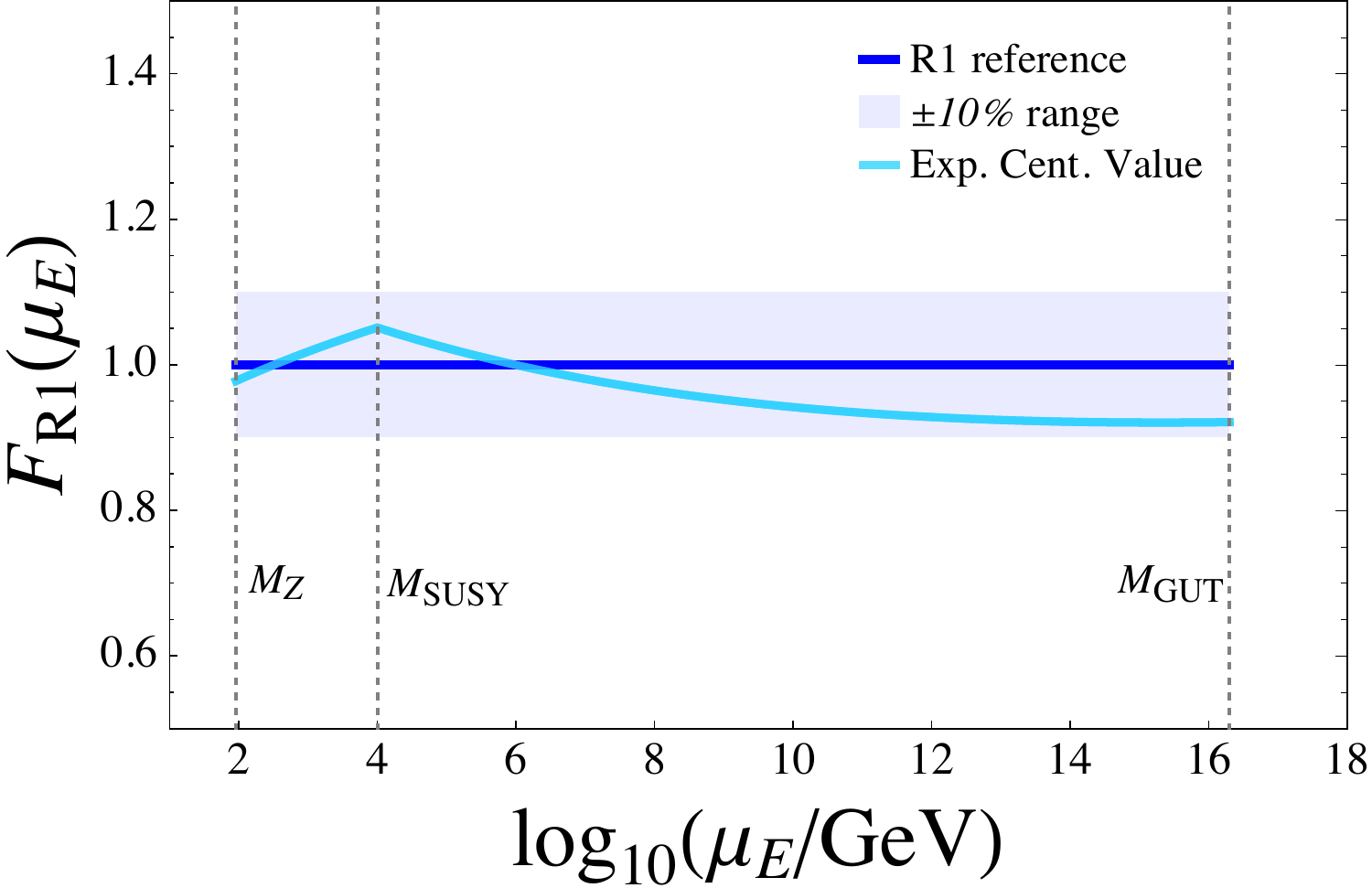} 
    \caption{$F_{\text{R1}}(\mu_E)$ from Eq. (\ref{eq:MassRelSqrd}) is plotted in cyan,
      with the values of the masses running from the electroweak scale $M_Z$ to the GUT scale $M_{\text{GUT}} = 2 \times 10^{16}$ GeV, using the REAP and SusyTC packages \cite{Antusch:2005gp,Antusch:2015nwi}.
      We take central values of quark and lepton masses \cite{Antusch:2013jca, Straub:2018kue}.
      The dark blue solid line $F_{\text{R1}}(\mu_E)=1$ is taken as reference and corresponds to the R1 mass relation. 
      The plot shows how its changes remain within the light-blue band corresponding to $10\%$ deviation, showing the potential validity of R1 from the electroweak to the GUT scale, given the uncertainty of the light quark masses $m_d$ and $m_s$.}
    \label{fig:MassRelation-Running}
  \end{figure}
  To quantify the RG running of the mass relation R1 we choose the mass of the $Z$-boson ($M_Z$) as the initial scale \cite{Antusch:2013jca,Straub:2018kue}, where it holds quite well, as seen
 from Figs. \ref{fig:MassRelation-Outer} and \ref{fig:MassRelation-Inner}.
For definiteness we choose as benchmark the following values for the SUSY breaking scale $M_{\text{SUSY}}=10$ TeV and $\tan{\beta}=\sfrac{v_u}{v_d}=50$.
Concerning the four CMSSM soft SUSY-breaking parameters, we take the common scalar mass $m_0=12$ TeV, fermion mass $m_{1/2}=10$ TeV, and the trilinear coupling $A_{0}=-15$ TeV. \par
With these assumptions, we plot in Fig.~\ref{fig:MassRelation-Running} the evolution of $F_{\text{R1}}(\mu_E)$ in Eq. (\ref{eq:MassRelSqrd}) from the electroweak scale $M_Z$ to the GUT scale
$M_{\text{GUT}} = 2 \times 10^{16}$ GeV, using the central values of the experimental quark and lepton masses as input.
We also display as a reference the case $F_{\text{R1}}(\mu_E)=1$, and the $\mathit{10} \%$ deviation band. 
It is clear from this figure that the mass relation is quite stable under RG running, and holds all the way from the electroweak scale to the GUT scale,
given the large uncertainty of the light quark masses $m_d$ and $m_s$. 
  
\section{Discussion and outlook }
\label{sec:discussion-outlook-}  
We have given a model-independent method to derive mass relations for the SM fermions in modular flavor symmetry models containing few parameters.
They are determined by the Clebsch-Gordan coefficients of the specific modular flavor group as well as the expansion coefficients of the corresponding vector-valued modular forms around the symmetry points.\par
We illustrated our results with a detailed example in Section \ref{Sec:ModularGamma4Explicit}, where we obtained viable mass relations for charged leptons and down-type quarks from the $\Gamma_4 \cong S_4$ flavor symmetry.
We showed that these mass relations are experimentally testable, and distinguishable from their non-modular counterpart the so-called golden mass relation given by
Eq.~(\ref{eq:MassRel2})~\cite{Morisi:2011pt,Bazzocchi:2012ve,King:2013hj,Bonilla:2014xla,Reig:2018ocz}. Since the mass relations can have sizable differences from one modular flavor group to another, they could be used to experimentally probe different models. \par 
In Appendix~\ref{sec:OtherMassRelations}  we point out the existence of a second class of mass relations that could yield viable predictions regarding neutrino masses and the up-quark sector.
Nonetheless, these are model-dependent. We presented a particular example to highlight the differences with respect to our general derivation in Section \ref{Sec:ModularGamma4Explicit}.
This example is very suggestive, as it relates the smallness of the solar squared mass splitting to the lightness of the up quark.\par
We want to stress that the examples presented in Section \ref{Sec:ModularGamma4Explicit} and Appendix~\ref{sec:OtherMassRelations} are not meant to be understood as complete flavor models,
rather as illustrations of the use of our general method for deriving mass relations. Note that, as discussed in the Introduction (Section \ref{sec:Introduction}), the main significance of the mass relations in
Eqs. (\ref{eq:MassRelMM})-(\ref{eq:MassRelPP}) and Eq.~(\ref{eq:MassRel2}) is that they are compatible with experimental data of the SM fermion masses, which is a non-trivial fact. 
Our two examples appear to be complementary, since in Section \ref{Sec:ModularGamma4Explicit} we only deal with the down sector of the MSSM, while in Appendix \ref{sec:OtherMassRelations} we discuss the up sector and neutrinos.
Since we present them independently, we can not make any meaningful statement concerning CKM mixing, which by definition involves both up and down quark sectors simultaneously.
Although one can complete each of the models trivially, we can not preserve both predictions simultaneously in a simple manner. 
In other words, it is straightforward to extend the model of the down sector in Section \ref{Sec:ModularGamma4Explicit} to include also the up-type quarks and neutrinos,
while preserving the mass relations  in Eqs. (\ref{eq:MassRelMM})-(\ref{eq:MassRelPP}), at the expense of losing the prediction for the up sector in Appendix \ref{sec:OtherMassRelations}. 
\par
It has been proven challenging to build experimentally viable quark and lepton modular flavor models while keeping the number of input parameters to a minimum. 
We expect that the results obtained in this work, largely model-independent, may prove useful also for more comprehensive modular invariant models and for top-down constructions. 
We demonstrated that viable mass relations for the SM fermions can emerge in modular flavor symmetry in a general manner
relying only on the modular flavor group and its vector-valued modular forms, rather than \textit{ad hoc} flavon alignments.
The mass relations can differentiate amongst models. This was illustrated with one explicit example where the mass relations predicted 
are distinguishable from the one obtained in a traditional flavor symmetry model with flavons.
The differences may be resolved experimentally and this may, perhaps, shed light on the symmetry underlying the flavor problem.  \par

\acknowledgements 
\noindent
We would like to thank  Stefan Antusch, Salvador Centelles Chuliá, Xueqi Li, Xiang-Gan Liu, and Michael Ratz for insightful discussions. SFK also thanks CERN for hospitality and Peter Stangl for very helpful discussions. We also thank Peter Stangl for providing us with the tools to significantly improve our plots using more updated values of light-quark masses.
OM thanks the Department of Physics and Astronomy at  UC, Irvine, for their hospitality during his visit.
SFK thanks IFIC, Valencia  for their hospitality during his visit where this work began.
Work supported by the Spanish grants PID2020-113775GB-I00 (AEI/10.13039/501100011033) and Prometeo CIPROM/2021/054 (Generalitat Valenciana).
OM acknowledges financial support from the Generalitat Valenciana through Programa Santiago Grisolía (No. GRISOLIA/2020/025) and the research visit grant CIBEFP/2022/63.
The research of MCC was supported, in part, by the U.S. National Science Foundation under Grant No. PHY-1915005.
SFK acknowledges the STFC Consolidated Grant ST/L000296/1 and the European Union's Horizon 2020 Research and Innovation programme under Marie Sklodowska-Curie grant agreement HIDDeN European ITN project (H2020-MSCA-ITN-2019//860881-HIDDeN).

\appendix

\section{The $\Gamma_4$ Group -  Basis and Modular Forms}
\label{Appendix:Gamma4Modular}

The $\Gamma_4 \simeq S_4$  group is of order $24$, all its elements can be written in terms of two generators $S$ and $T$ following its presentation equation:
\begin{equation}
S_4\simeq \{S,T|S^2=T^4=(ST)^3=e\}.
\end{equation}
The $S_4$ group has five irreducible representations: two singlets $\bf{1}$, $\bf{1^{\prime}}$, one doublet  $\bf{2}$, and two triplets $\bf{3}$, $\bf{3}^{\prime}$.
In this work we use the basis displayed in Table \ref{tab:Gamma4GenRep}.
In this basis the $T$ generator is diagonal for all irreducible representations, and both generators are symmetric matrices. \par  
\begin{table}[h]
\centering
\begin{tabular}{|c|c|c|}\hline\hline
 & $S$ & $T$ \\ \hline
  $\mathbf{1},\mathbf{1^{\prime}}$ & $\pm 1$ & $\pm 1$  \\ \hline
  $\mathbf{2}$ & $\dfrac{1}{2}\begin{pmatrix}
  -1 & \sqrt{3} \\
 \sqrt{3} & 1 \\
\end{pmatrix}$ & $\begin{pmatrix}
 1 & 0 \\
 0 & -1 \\
\end{pmatrix}$  \\ \hline
    $\mathbf{3},\mathbf{3^{\prime}}$ & $\pm\dfrac{1}{2}\begin{pmatrix}
 0 & \sqrt{2} & \sqrt{2} \\
 \sqrt{2} & -1 & 1 \\
 \sqrt{2} & 1 & -1 \\
\end{pmatrix}$ & $\pm\begin{pmatrix}
 1 & 0 & 0 \\
 0 & i & 0 \\
 0 & 0 & -i \\
\end{pmatrix}$  \\ \hline
\end{tabular}
\caption{Generators of $S_{4}$ in the symmetric $T$-diagonal basis. }
\label{tab:Gamma4GenRep}
\end{table}

The $S_4$ representation products relevant for the examples outlined in Section \ref{Sec:ModularGamma4Explicit} and Appendix \ref{sec:OtherMassRelations} are given by
\begin{align}
\begin{array}{@{}cll@{}}
\mathbf{3}\,\otimes\,\mathbf{3'}\, \Longrightarrow &
\left\{\begin{array}{@{}l@{\hspace{3pt}\sim\hspace{3pt}}l@{}}
\hspace{3.5pt} \mathbf{1'} & \alpha_1\beta_1+\alpha_2\beta_3+\alpha_3\beta_2\\[2mm]
\hspace{3.5pt} \mathbf{2}  & \begin{pmatrix}
                      \frac{\sqrt{3}}{2}\left(\alpha_2\beta_2+\alpha_3\beta_3\right)  \\
                      -\alpha_1\beta_1+ \frac{1}{2}\left(\alpha_2\beta_3+\alpha_3\beta_2\right)                   
                    \end{pmatrix}\\[4mm]
\hspace{3.5pt} \mathbf{3}   &\begin{pmatrix}
                       \alpha_3\beta_3-\alpha_2\beta_2 \\
                       \alpha_1\beta_3+\alpha_3\beta_1 \\
                       -\alpha_1\beta_2-\alpha_2\beta_1
                     \end{pmatrix}\\[6mm]
\hspace{3.5pt} \mathbf{3'}  & \begin{pmatrix} 
                       \alpha_3\beta_2-\alpha_2\beta_3 \\
                       \alpha_2\beta_1-\alpha_1\beta_2\\
                       \alpha_1\beta_3-\alpha_3\beta_1
                     \end{pmatrix} 
\end{array}\right.
\end{array} \quad 
\begin{array}{@{}cll@{}}
\mathbf{3}\,\otimes\,\mathbf{3}\Longrightarrow &
\left\{\begin{array}{@{}l@{\hspace{3.5pt}\sim\hspace{3.5pt}}l@{}}
\hspace{3.5pt} \mathbf{1}  & \alpha_1\beta_1+\alpha_2\beta_3+\alpha_3\beta_2\\[2mm]
\hspace{3.5pt} \mathbf{2}  & \begin{pmatrix}
                      \alpha_1\beta_1- \frac{1}{2}\left(\alpha_2\beta_3+\alpha_3\beta_2\right) \\
                    \frac{\sqrt{3}}{2} \left(\alpha_2\beta_2+\alpha_3\beta_3\right) 
                    \end{pmatrix}\\[4mm]
\hspace{3.5pt} \mathbf{3}   &\begin{pmatrix}
                       \alpha_3\beta_2-\alpha_2\beta_3 \\
                       \alpha_2\beta_1-\alpha_1\beta_2\\
                       \alpha_1\beta_3-\alpha_3\beta_1
                     \end{pmatrix} \\[6mm]
\hspace{3.5pt} \mathbf{3'}  & \begin{pmatrix} 
                       \alpha_3\beta_3-\alpha_2\beta_2 \\
                       \alpha_1\beta_3+\alpha_3\beta_1 \\
                       -\alpha_1\beta_2-\alpha_2\beta_1
                     \end{pmatrix}
\end{array}\right.
\end{array}
\end{align}
\begin{align}
\begin{array}{@{}c@{{}\,\otimes\,{}}c@{{}\,\,\,{}}ll@{}}
\mathbf{2}&\mathbf{2}& \Longrightarrow &
\left\{\begin{array}{@{}l@{\hspace{3.5pt}\sim\hspace{3.5pt}}l@{}}
\hspace{3.5pt} \mathbf{1}  & \alpha_1\beta_1+\alpha_2\beta_2\\[2mm]
\hspace{3.5pt} \mathbf{1'} & \alpha_1\beta_2-\alpha_2\beta_1\\[2mm]
\hspace{3.5pt} \mathbf{2}  & \begin{pmatrix} \alpha_2\,\beta_2 - \alpha_1\,\beta_1\\
                                    \alpha_1\,\beta_2 + \alpha_2\,\beta_1
                    \end{pmatrix}
\end{array}\right.
\end{array}
\end{align}
The vector-valued modular forms of the $\Gamma_4$ group at weight $2$ are 
\begin{equation}
Y^{(2)}_{\mathbf{2}}(\tau)=\begin{pmatrix}Y_1(\tau)\\ Y_2(\tau)\end{pmatrix},\hspace{10pt} Y^{(2)}_{\mathbf{3}}(\tau)=\begin{pmatrix}Y_3(\tau)\\ Y_4(\tau)\\  Y_5(\tau)\end{pmatrix}.
\label{eq:S4ModularFormsWeight2Appendix}
\end{equation} 
these can be written as power expansions
\begin{equation}
Y_{i}(\tau) = \sum^{\infty}_{n=0} c_{i,n} q^{\frac{n}{4}}\hspace{2pt},\qquad \text{with}\qquad  q\equiv e^{2\pi i \tau},
\label{eq:YExp}
\end{equation}
where $c_{i,n}$ are constant coefficients. We obtained the explicit expansion following the derivation in \cite{Ding:2022nzn}.
\begin{align}
Y_{1}(\tau) &= 1 + 24q +24 q^2+96 q^3+24 q^4+144 q^5+96 q^6+192 q^7+...\hspace{2pt}, \nonumber \\
Y_{2}(\tau) &= q^{\frac{1}{2}} (-8 \sqrt{3}  - 32 \sqrt{3} q - 48\sqrt{3} q^2 - 64\sqrt{3} q^3 - 104\sqrt{3} q^4  - 96 \sqrt{3} q^5+ ...\hspace{2pt}), \nonumber \\
Y_{3}(\tau) &= 1 - 8 q + 24 q^2 - 32 q^3 + 24 q^4 - 48 q^5 + 96 q^6 - 64 q^7+...\hspace{2pt}, \label{eq:YExpEpsilon}\\
Y_{4}(\tau) &= q^{\frac{1}{4}}(-4 \sqrt{2} - 24 \sqrt{2} q - 52 \sqrt{2} q^2 - 
 56 \sqrt{2} q^3 - 72 \sqrt{2} q^4 - 128 \sqrt{2} q^5)...\hspace{2pt}, \nonumber \\
Y_{5}(\tau) &= q^{\frac{3}{4}}(-16 \sqrt{2} - 32 \sqrt{2} q^2 - 48 \sqrt{2} q^3 - 
 96 \sqrt{2} q^4 - 80 \sqrt{2} q^5 +...)\hspace{2pt}. \nonumber 
\end{align}

\section{Polynomial system solutions}  
\label{Appendix:Polynomials}

In Eqs. (\ref{eq:TrHSym}) and (\ref{eq:Tr2HSym}) of Section \ref{sec:MassRelations} we defined the polynomials $h_\psi(a_1,a_2)$ and $g_{\psi}(a_1,a_2)$,
these are homogeneous polynomials of order $2$ and $4$ respectively. Ultimately the solutions of the polynomial system 
\begin{align}
\begin{split}
h_\psi(a_1,a_2) &= m^2_2 + m^2_3, \\
g_\psi(a_1,a_2) &= m^2_2\, m^2_3, \label{eq:PolynomialSystem}
\end{split}
\end{align}
determine $f(m_3,m_2)$ in Eq. (\ref{eq:MassesAndEpsilon}), which is the main result of this work.\par 
By naive counting one could infer that this polynomial system has eight solutions, the product of the orders of the two polynomials. 
Nonetheless the actual number of independent and physically inequivalent solutions of the polynomial system in (\ref{eq:PolynomialSystem})
cannot be determined generically without the explicit form of  $h_\psi(a_1,a_2)$  and $g_{\psi}(a_1,a_2)$.\par 
To illustrate this, consider the mass matrix $M_f$ in Eq. (\ref{eq:MdAndMeMatrix}), for this particular case the polynomials take the form
\begin{equation}
h_\psi(a_1, a_2) = \frac{3}{2}{a_1}^2 + 2 {a_2}^2 = m^2_2 + m^2_3, \qquad g_\psi(a_1, a_2) = \frac{1}{16} \left( 3 {a_1}^2 -4 {a_2}^2\right)^2= m^2_2\, m^2_3.
\label{eq:HfPoly}
\end{equation}
This polynomial system has in fact only three independent and physically inequivalent solutions  
\begin{align}
\textbf{a)}\quad\qquad \tilde{a}_{1,+}(m_2,m_3) &= \frac{m_3+ m_2}{\sqrt{3}}, \qquad \tilde{a}_{2,-}(m_2,m_3) = \frac{m_3 - m_2}{2},
\label{eq:SolsI}\\
\textbf{b)}\quad\qquad\tilde{a}_{1,-}(m_2,m_3) &= \frac{m_3 - m_2}{\sqrt{3}}, \qquad \tilde{a}_{2,+}(m_2,m_3) = \frac{m_3+ m_2}{2},
\label{eq:SolsII}\\
\textbf{c)}\quad\qquad \hat{a}_{1,+}(m_2,m_3) &=  \frac{m_3 + m_2}{\sqrt{3}}, \qquad \hat{a}_{2,-}(m_2,m_3) = -\frac{m_3- m_2}{2},
\label{eq:SolsIII}
\end{align} 
These solutions, following Eqs. (\ref{eq:fpolyMasses}) and (\ref{eq:MassesAndEpsilon}) lead to the following three mass relations around the $\tau_{\text{sym}}=i\infty$,
in terms of $\epsilon$ defined in Eq. (\ref{eq:epsilon})
\begin{equation}
\textbf{a)} \quad \frac{m_2 m_1}{8 m_3(m_3-3m_2)} \approx \lvert\epsilon\rvert^2\hspace{2pt},\qquad\quad \textbf{b)} \quad \frac{m_2 m_1}{8 m_3(m_3+3m_2)} \approx \lvert\epsilon\rvert^2\hspace{2pt},\qquad\quad
\textbf{c)}\quad \frac{m_3 m_1}{8 m_2(m_2+3m_3)} \approx \lvert\epsilon\rvert^2\hspace{2pt}.
\label{eq:AllMassRelations}
\end{equation}
Solutions \textbf{a)} and \textbf{b)} yield viable predictions when relating down-type quark and charged-lepton masses, as they feature a leading power of $m^2_3$ in the denominator (refer to Subsection \ref{sec:zooming-golden-quark} for more details).
 In a complementary fashion, solution \textbf{c)} provides novel viable predictions for the up-sector masses, as demonstrated by Eq. (\ref{eq:UpSectorMassRel}) in Appendix \ref{sec:OtherMassRelations}. 

\section{Another class of mass relations}
\label{sec:OtherMassRelations}

Our derivation in Section \ref{sec:MassRelations} is general, and model-independent.
We now consider a second class of mass relations that can emerge in modular invariant models.
This happens when an $H_{\psi}$ matrix, defined in Eq. (\ref{eq:Hpsi}), satisfies the first condition listed in Subsection \ref{subsec:conditions},
while it does not satisfy the second. Meaning that at the symmetric point $\epsilon \to 0 $ the rank of the matrix is not reduced.  
\begin{equation}
\text{ rank} \left[ \lim_{\epsilon \to 0} H_{\psi}(a_1,a_2,\epsilon)\right] = \text{rank}[H_{\psi}(a_1,a_2,\epsilon)].
\label{eq:MNotRankReduction}
\end{equation}
This implies that, in the limit, the three non-vanishing masses $m_1$, $m_2$ and $m_3$ can be written as functions of only two parameters $a_1$, and $a_2$.
This yields algebraic relations amongst the three masses at the symmetry point.  
This relation can be obtained by solving the polynomial system in Eqs. (\ref{eq:TrHSym}), and (\ref{eq:Tr2HSym}).
However, it cannot be derived in a model independent way, as it does not only depend on the representation and weight of the fields, but also on the specific symmetry point.

\subsection{Example of a mass relation for the up-type quarks and neutrinos} 
\label{sec:example-mass-relat}

We regard it illustrative to give an explicit example of how to obtain the second kind of mass relations which could also potentially yield interesting and viable predictions. \par 
\begin{table}[h]
\centering
\begin{tabular}{|c|c|c|c|c|c|c|c|}
\hline
\textbf{}      & $Q$          & $U^c$        & $L$          & $N^c$        & $\Phi_u$        & $Y^{(2)}_{\mathbf{2}}$ & $Y^{(2)}_{\mathbf{3}}$ \\ \hline
\textbf{$S_4$} & $\mathbf{3}$ & $\mathbf{3^{\prime}}$ & $\mathbf{3}$ & $\mathbf{3}$ & $\mathbf{1}$ & $\mathbf{2}$           & $\mathbf{3}$           \\ \hline
\textbf{$k$}   & $-1$         & $-1$         & $-2$         & $0$         & $0$         & $2$                    & $2$                    \\ \hline
\end{tabular}
\caption{
  $\Gamma_4$ weight and representation assignments that lead to a correlation between neutrino and up-quark masses. Gauge MSSM transformation properties are omitted.}
\label{tab:S4IrrepExampleUpSector}
\end{table}\par  
For definiteness, we assume that neutrinos acquire a mass through a Type-I seesaw mechanism, including thus the $N^c$ superfields for the three heavy right-handed (RH) neutrinos.
The representation and weight assignments are given in Table \ref{tab:S4IrrepExampleUpSector}, leading to the following up-sector MSSM superpotential terms  
\begin{align}
\mathcal{W}^{\Gamma_4}_{\Phi_u} &\supset \hspace{3pt} \alpha^{u}_{1} \left(Q \Phi_u U^c Y^{(2)}_{\mathbf{2}}(\tau) \right)_{\mathbf{1}} + \hspace{3pt}  \alpha^{u}_{2} \left(Q \Phi_u U^c Y^{(2)}_{\mathbf{3}}(\tau) \right)_{\mathbf{1}} \nonumber\\
 &\hspace{4pt}+\alpha^{\nu}_{1} \left(L \Phi_u N^c Y^{(2)}_{\mathbf{2}}(\tau) \right)_{\mathbf{1}} + \hspace{3pt} \alpha^{\nu}_{2} \left(L \Phi_{u} N^c Y^{(2)}_{\mathbf{3}}(\tau) \right)_{\mathbf{1}} + \frac{m_{N}}{2} N^c N^c.
\label{eq:YukawaS4Up}
\end{align}
Here the modular forms are the ones given in Eq. (\ref{eq:S4ModularFormsWeight2}). The Dirac and RH neutrino mass blocks are given by 
\begin{equation}
M^D_{\nu} = 
\begin{pmatrix}
a^{\nu}_{1} Y_{1}  & -a^{\nu}_{2} Y_{5} & a^{\nu}_{2} Y_{4} \\
a^{\nu}_{2} Y_{5} & \frac{\sqrt{3}}{2}  a^{f}_{1} Y_{2}  & -\frac{1}{2} \left(a^{\nu}_{1} Y_{1} + 2 a^{\nu}_{2} Y_3 \right) \\
-a^{f}_{2} Y_{4}  & -\frac{1}{2} \left(a^{\nu}_{1} Y_{1} - 2 a^{\nu}_{2} Y_3 \right) &  \frac{\sqrt{3}}{2} a^{\nu}_{1} Y_2 \\
\end{pmatrix}, \qquad  \text{and}, \qquad 
M_{N}  = m_N
\begin{pmatrix}
1&0&0\\
0&0&1\\
0&1&0
\end{pmatrix}.
\label{eq:NeutrinoMassMatrices}
\end{equation}
The effective mass of the light neutrinos is given by the standard seesaw formula 
\begin{equation}
M_\nu \approx -M^D_{\nu} M^{-1}_N ({M^D_\nu})^\top.
\label{eq:Seesaw}
\end{equation}
We choose these particular representations to illustrate a relevant point. The Hermitian matrix $H_{\nu}\equiv M_\nu M^\dagger_{\nu}$ fulfills the first property in
subsection \ref{subsec:conditions}, with only two dimensionful parameters; $a^{\nu}_1$, and $a^{\nu}_2$. However, at the symmetry point $\tau_{\text{sym}} = i \infty$ its
rank is not reduced, Eq.  (\ref{eq:MNotRankReduction}). \par 
In this example, for simplicity we choose the up-quark mass matrix $M_u$ to have the same structure that $M_f$ in Eq. (\ref{eq:MdAndMeMatrix}) thus around $\tau_{\text{sym}} = i \infty$
we have the following expression relating up-quark masses 
\begin{equation}
\frac{m^2_t m^2_{c} m^2_u}{f(m_t,m_c)}\approx \lvert\epsilon\rvert^4
\label{eq:upquarkDep}
\end{equation}
where the polynomial can be either one of the three solutions listed in Eq. (\ref{eq:AllMassRelations}) in Appendix \ref{Appendix:Polynomials}. 
The last equation indicates that at this symmetric point the masses of the three families satisfy 
\begin{equation}
\lim_{\epsilon \to 0 }  \begin{pmatrix} m_t \\ m_c \\ m_u \end{pmatrix} = \begin{pmatrix} m_t\left(a^u_1,a^u_2\right)\\ m_c\left(a^u_1,a^u_2\right) \\ 0 \end{pmatrix}, \qquad \quad \lim_{\epsilon \to 0 }  \begin{pmatrix} m^{\nu}_3 \\ m^{\nu}_2 \\ m^{\nu}_1 \end{pmatrix} = \begin{pmatrix} m^{\nu}_3 \left(a^\nu_1,a^\nu_2\right)\\ m^{\nu}_2\left(a^\nu_1,a^\nu_2\right) \\ m^{\nu}_1 \left(a^\nu_1,a^\nu_2\right) \end{pmatrix},
\label{eq:FixedPointMases}
\end{equation}
Notice that $m_u$ vanishes at the symmetric limit as expected from the relation in Eq. (\ref{eq:upquarkDep}). 
However, since neither of the neutrino masses vanish, the three neutrino masses are not independent. In fact at the symmetry point two neutrino masses are degenerate: 
\begin{equation}
\lim_{\tau \to i \infty}  m^{\nu}_2 = m^{\nu}_1\qquad  \Longrightarrow \quad  \lim_{\tau \to i \infty}  \Delta m^2_{21} = 0.
\label{eq:symLimNuMass}
\end{equation}
Therefore in this example the smallest of the required neutrino mass squared splitting, the solar $\Delta m^{2}_{21}$, must result from a departure from the symmetry point $\tau_T = i \infty$ yielding $\Delta m^2_{21} \propto \lvert \epsilon \rvert^2$. \footnote{Note that in this example the relation between the three neutrino masses depends on the value of $\epsilon(\tau)$, in contrast to the one obtained in reference \cite{CentellesChulia:2023zhu}.}
For this illustrative example we restrict ourselves to the region Im $ \epsilon  = 0 $, hence for values $ \lvert \epsilon \rvert \ll 1$ we obtain the simplified expression
\begin{equation}
\Delta m^2_{21}  \approx f^{\nu}_{\pm}(m^\nu_3, m^\nu_1) \lvert \epsilon \rvert^2,
\label{eq:SolarSymDev}
\end{equation}
This expression is analogous to Eq. (\ref{eq:MassesAndEpsilon}) but manifestly different. The function $f^{\nu}_{\pm}( m^\nu_3, m^\nu_1)$ has mass-dimension $2$ and is given by 
\begin{equation} 
f^{\nu}_{\pm}( m^\nu_3, m^\nu_1) = \frac{8 m^{\nu}_1}{\Delta m^{2}_{31}}\left[8 {m^{\nu}_1}^3  + 16 {m^{\nu}_1}^2 {m^{\nu}_3}+ 14 {m^{\nu}_1} {m^{\nu}_3}^2 - 3{m^{\nu}_3}^3 \pm 3\sqrt{4 {m^{\nu}_1} + {m^{\nu}_3}} \left( 4{m^{\nu}_1} {m^{\nu}_3}^{\frac{3}{2}}  + {m^{\nu}_3}^{\frac{5}{2}} \right)  \right],
\label{eq:NeutrinoSectorFunction}
\end{equation}
where $\Delta m^2_{31}$ is the atmospheric neutrino squared mass difference.
 Notice that last equation is valid for {both normal (NO) and inverted ordering (IO) of neutrino masses since the polynomial form is common.\par 
A viable correlation between the up-quark and neutrino masses can be obtained by using Eqs. (\ref{eq:upquarkDep}), (\ref{eq:NeutrinoSectorFunction}), the plus sign function $f^{\nu}_+$ and the third solution in Eq. (\ref{eq:AllMassRelations}), i.e.
\begin{equation}
\frac{\Delta m^2_{21} }{f^{\nu}_{+}(m^\nu_3, m^\nu_1)} \approx \frac{m_t m_{c} m_u}{8m^2_c(m_c+3m_t)}
\label{eq:UpSectorMassRel}.
\end{equation}
This correlation becomes manifest when we scan our parameter space $\{a^{u,\nu}_{1,2}, \lvert \epsilon \rvert \} \Longrightarrow \{m_{u,c,t},m^{\nu}_{1,2,3} \}$ as shown in Fig. \ref{fig:MassRelation-Up}. 
\begin{figure}[!t]
    \centering
    \begin{subfigure}{0.46\textwidth}
        \includegraphics[width=\textwidth]{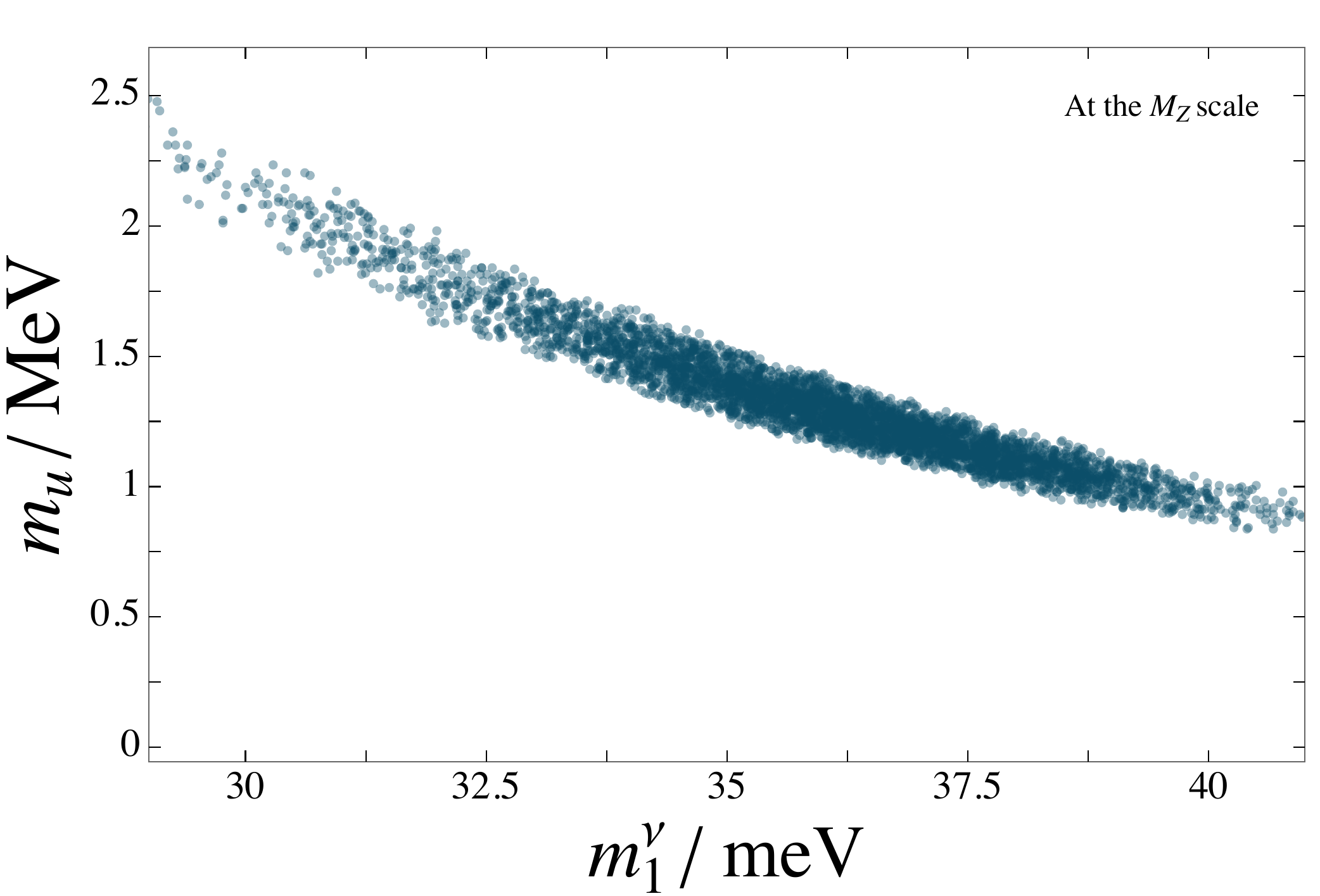}
        \label{fig:sub1}
    \end{subfigure}
    \hfill 
    \begin{subfigure}{0.47\textwidth}
        \includegraphics[width=\textwidth]{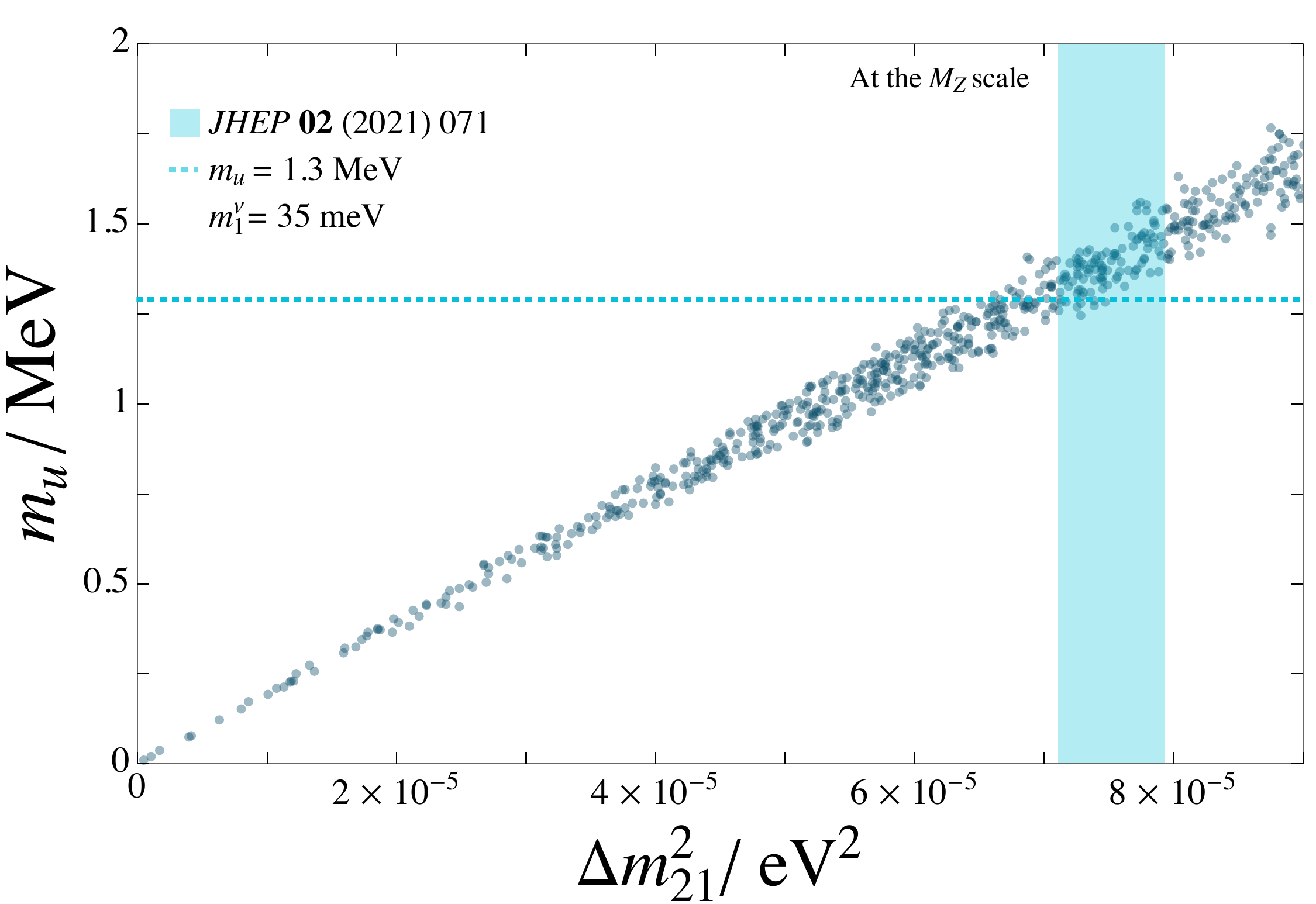}
        \label{fig:sub2}
    \end{subfigure}
    \caption{ Predicted quark-lepton mass correlations. 
        In the left panel we show $m_u$ versus the lightest neutrino mass $m^{\nu}_1$, each point lies inside the experimental $3$-$\sigma$ range for $m_t$, $m_c$, $\Delta m^2_{21}$, and $\Delta m^2_{31}$ from PDG~\cite{Workman:2022ynf} and~\cite{deSalas:2020pgw} respectively.
        This correlation is the prediction in Eq. (\ref{eq:UpSectorMassRel}).
        The right panel assumes normal ordering, fixing $m^{\nu}_1 = 35$ meV.
        It illustrates how both $m_u$ and $\Delta m^2_{21}$ are generated simultaneously, when departing from the symmetry point as predicted by Eqs. (\ref{eq:upquarkDep}) and (\ref{eq:SolarSymDev}).}
    \label{fig:MassRelation-Up}
\end{figure}
Indeed, this figure shows a correlation in the $m_u$ vs. squared solar mass splitting plane that results by taking 
points $\{m_t,\hspace{1pt} m_c, \hspace{1pt} \Delta m^2_{21}, \hspace{1pt} \Delta m^2_{31} \text{ (NO)}  \}$  within their
$3\sigma$ ranges given the PDG~\cite{Workman:2022ynf} and Ref.~\cite{deSalas:2020pgw}, respectively.
This result is very suggestive, as it correlates the smallness of the solar squared mass splitting to the lightness of $m_u$, which is the quark mass with the largest uncertainty \cite{Antusch:2013jca}.\par 
This example is relevant as it shows that there is a (second) kind of mass relations that can emerge in modular symmetric models that is not included in our general derivation in Section \ref{sec:MassRelations}.  
We want to highlight that this example, and in particular Eqs. (\ref{eq:symLimNuMass}) and (\ref{eq:SolarSymDev}), illustrates how mass matrices satisfying Eq. (\ref{eq:MNotRankReduction})
could lead to viable predictions for the neutrino sector. 

\newpage

\bibliographystyle{utphys}
\bibliography{bibliography}
\end{document}